\documentclass[12pt]{article}

\makeatletter

\def\paragraph{\@startsection{paragraph}{4}{\z@}{+2.00ex plus
 +1ex minus +.2ex}{1.5ex plus .2ex}{\it\normalsize}}

\def\section{\@startsection {section}{1}{\z@}{+3.0ex plus +1ex minus
  +.2ex}{2.3ex plus .2ex}{\normalsize\bf\boldmath}}
\def\subsection{\@startsection{subsection}{2}{\z@}{+2.5ex plus +1ex
minus +.2ex}{1.5ex plus .2ex}{\normalsize\bf\boldmath}}
\def\subsubsection{\@startsection{subsubsection}{3}{\z@}{+3.25ex plus
 +1ex minus +.2ex}{1.5ex plus .2ex}{\normalsize\it}}

\expandafter\ifx\csname mathrm\endcsname\relax\def\mathrm#1{{\rm #1}}\fi

\newcounter{saveeqn}

\@addtoreset{equation}{section}

\newcount\@tempcntc
\def\@citex[#1]#2{\if@filesw\immediate\write\@auxout{\string\citation{#2}}\fi
  \@tempcnta\z@\@tempcntb\m@ne\def\@citea{}\@cite{\@for\@citeb:=#2\do
    {\@ifundefined
       {b@\@citeb}{\@citeo\@tempcntb\m@ne\@citea
        \def\@citea{,\penalty\@m\ }{\bf ?}\@warning
       {Citation `\@citeb' on page \thepage \space undefined}}%
    {\setbox\z@\hbox{\global\@tempcntc0\csname
b@\@citeb\endcsname\relax}%
     \ifnum\@tempcntc=\z@ \@citeo\@tempcntb\m@ne
       \@citea\def\@citea{,\penalty\@m}
       \hbox{\csname b@\@citeb\endcsname}%
     \else
      \advance\@tempcntb\@ne
      \ifnum\@tempcntb=\@tempcntc
      \else\advance\@tempcntb\m@ne\@citeo
      \@tempcnta\@tempcntc\@tempcntb\@tempcntc\fi\fi}}\@citeo}{#1}}

\def\@citeo{\ifnum\@tempcnta>\@tempcntb\else\@citea
  \def\@citea{,\penalty\@m}%
  \ifnum\@tempcnta=\@tempcntb\the\@tempcnta\else
   {\advance\@tempcnta\@ne\ifnum\@tempcnta=\@tempcntb \else
\def\@citea{--}\fi
    \advance\@tempcnta\m@ne\the\@tempcnta\@citea\the\@tempcntb}\fi\fi}

\def\nl{\nonumber\\}
\def\nls{\nonumber\\[1ex]}
\def\nnls{\nonumber\\[1ex]}

\def\nln{\nl[-1ex]}

\def\asymp#1%
{\mathrel{\raisebox{-.4em}{$\widetilde{\scriptstyle #1}$}}}

\def\Nequal#1%
{\mathrel{\raisebox{-.5em}{$\stackrel{=}{\scriptstyle\rm#1}$}}}
\newcommand{\dsl}[1]{\not \hspace{-0.7mm}#1}
\def\dsl{\mathpalette\make@slash}
\def\make@slash#1#2{\setbox\z@\hbox{$#1#2$}%
  \hbox to 0pt{\hss$#1/$\hss\kern-\wd0}\box0}

\newcommand{\asymLa}{\asymp{\La\to\infty}}

\def\beq{\begin{equation}}
\def\eeq{\end{equation}}
\def\beqar{\begin{eqnarray}}
\def\eeqar{\end{eqnarray}}
\def\barr#1{\begin{array}{#1}}
\def\earr{\end{array}}
\def\bfi{\begin{figure}}
\def\efi{\end{figure}}
\def\btab{\begin{table}}
\def\etab{\end{table}}
\def\bce{\begin{center}}
\def\ece{\end{center}}
\def\nn{\nonumber}
\def\disp{\displaystyle}
\def\text{\textstyle}

\def\al{\alpha}

\def\de{\delta}
\def\De{\Delta}

\def\La{\Lambda}

\def\si{\sigma}

\def\refeq#1{\mbox{(\ref{#1})}}

\def\refse#1{\mbox{Section~\ref{#1}}}

\def\refapp#1{\mbox{App.~\ref{#1}}}
\def\citere#1{\mbox{Ref.~\cite{#1}}}
\def\citeres#1{\mbox{Refs.~\cite{#1}}}

\newcommand{\ri}{{\mathrm{i}}}
\newcommand{\rd}{{\mathrm{d}}}

\renewcommand{\O}{\mathswitch{{\cal{O}}}}

\def\mathswitchr#1{\relax\ifmmode{\mathrm{#1}}\else$\mathrm{#1}$\fi}
\newcommand{\Pf}{\mathswitch  f}

\newcommand{\PH}{\mathswitchr H}

\newcommand{\Pt}{\mathswitchr t}

\newcommand{\Pep}{\mathswitchr {e^+}}
\newcommand{\Pem}{\mathswitchr {e^-}}

\def\mathswitch#1{\relax\ifmmode#1\else$#1$\fi}

\def\ie{i.e.,\ }
\def\eg{e.g.\ }

\newcommand{\textfrac}[2]{{\textstyle\frac{#1}{#2}}}
\newcommand{\D}{{\cal{D}}}
\newcommand{\Btilde}{{\tilde{B}}}
\newcommand{\Ctilde}{{\tilde{C}}}
\newcommand{\Dtilde}{{\tilde{D}}}
\newcommand{\Ntilde}{{\tilde{N}}}
\newcommand{\extradet}{V}
\newcommand{\UVdet}{U}
\newcommand{\Zhat}{\hat Z}

\newcommand{\debar}{\bar\delta}
\newcommand{\detY}[1]{\eta_{#1}}
\newcommand{\detZh}[1]{\zeta_{#1}}
\newcommand{\Dcomb}{D}
\newcommand{\Ccomb}{C}
\newcommand{\Bcomb}{B}
\newcommand{\fin}{\mathrm{(fin)}}

\def\draftdate{\relax}
\def\mda{\relax}
\def\mua{\relax}
\def\mla{\relax}
\def\draft{
\def\thtystars{******************************}
\def\sixtystars{\thtystars\thtystars}
\typeout{}
\typeout{\sixtystars**}
\typeout{* Draft mode!
         For final version remove \protect\draft\space in source file *}
\typeout{\sixtystars**}
\typeout{}
\def\draftdate{\today}
\def\mua{\marginpar[\boldmath\hfil$\uparrow$]%
                   {\boldmath$\uparrow$\hfil}%
                    \typeout{marginpar: $\uparrow$}\ignorespaces}
\def\mda{\marginpar[\boldmath\hfil$\downarrow$]%
                   {\boldmath$\downarrow$\hfil}%
                    \typeout{marginpar: $\downarrow$}\ignorespaces}
\def\mla{\marginpar[\boldmath\hfil$\rightarrow$]%
                   {\boldmath$\leftarrow $\hfil}%
                    \typeout{marginpar: $\leftrightarrow$}\ignorespaces}
\def\Mua{\marginpar[\boldmath\hfil$\Uparrow$]%
                   {\boldmath$\Uparrow$\hfil}%
                    \typeout{marginpar: $\uparrow$}\ignorespaces}
\def\Mda{\marginpar[\boldmath\hfil$\Downarrow$]%
                   {\boldmath$\Downarrow$\hfil}%
                    \typeout{marginpar: $\downarrow$}\ignorespaces}
\def\Mla{\marginpar[\boldmath\hfil$\Rightarrow$]%
                   {\boldmath$\Leftarrow $\hfil}%
                    \typeout{marginpar: $\leftrightarrow$}\ignorespaces}
\overfullrule 5pt
\oddsidemargin -15mm
\marginparwidth 29mm
}

\def\stars{\strut\leaders\hbox{*}\hfill\strut}
\def\starline{\hfil\strut\hfil\hbox to \textwidth {\stars}\hfil}

\begin{document}
\thispagestyle{empty}
\def\thefootnote{\fnsymbol{footnote}}
\setcounter{footnote}{1}
\null
\draftdate
\strut\hfill MPI-PhT/2002-63\\
\strut\hfill PSI-PR-02-21\\
\strut\hfill hep-ph/0212259 
\vfill
\begin{center}
{\large \bf\boldmath
Reduction of one-loop tensor 5-point integrals
\par} \vskip 2em
\vspace{1cm}
{\large
{\sc A.\ Denner$^1$ and S.\ Dittmaier$^2$} } 
\\[.5cm]
$^1$ {\it Paul Scherrer Institut\\
CH-5232 Villigen PSI, Switzerland} 
\\[0.3cm]
$^2$ {\it  Max-Planck-Institut f\"ur Physik (Werner-Heisenberg-Institut) \\
             F\"ohringer Ring 6, D-80805 M\"unchen, Germany}
\par 
\end{center}\par
\vfill
\vskip 2.0cm {\bf Abstract:} \par 
A new method for the reduction of one-loop tensor 5-point integrals to
related 4-point integrals is proposed.  In contrast to the usual
Passarino--Veltman reduction and other methods used in the literature,
this reduction avoids the occurrence of inverse Gram determinants,
which potentially cause severe numerical instabilities in 
practical calculations. Explicit results for the 5-point
tensor coefficients are presented up to rank 4.  
The expressions for the reduction of the relevant 1-, 2-, 3-, and
4-point tensor coefficients to scalar integrals are also included;
apart from these standard integrals no other integrals are needed.

\par
\vskip 1cm
\noindent
December 2002   
\null
\setcounter{page}{0}
\clearpage
\def\thefootnote{\arabic{footnote}}
\setcounter{footnote}{0}

\section{Introduction}
\label{se:intro}

High-energy collider experiments reached a 
high level of accuracy
in the last decades. For instance, the LEP, SLC, and Tevatron experiments 
tested the Standard Model of electroweak and strong interactions as 
quantum field theories, in the sense that quantum corrections, \ie
higher-order perturbative radiative corrections, had to be taken into
account to successfully compare predictions with data. The most precise
results typically were obtained from investigations of $1\to2$ particle
decays and $2\to2$ scattering reactions. At present and future colliders,
such as the Tevatron, the LHC, and an $\Pep\Pem$ linear collider,
scattering reactions with more final-state particles will gain increasing 
interest in various contexts, such as in the production of jets, 
heavy quark flavours, electroweak gauge bosons, and Higgs bosons.

In this paper we focus on $2\to3$ particle reactions and present one
of the basic ingredients in the evaluation of radiative corrections
at the one-loop level, the calculation of 5-point tensor integrals.
Following the well-known procedure of Passarino and Veltman
\cite{Passarino:1978jh}, one-loop tensor integrals can be recursively
reduced to scalar integrals by solving sets of linear equations for
the tensor coefficients in the most general Lorentz-covariant 
ansatz.
Specifically, for each tensor rank this procedure involves a factor 
of the inverse Gram matrix which is built by the momenta that span
the tensor. Since these momenta become linearly dependent at the
boundary of phase space, the inverse Gram matrix becomes singular 
at the boundary. In practice, this leads to numerical instabilities
near the phase-space boundary, where the zero in the Gram determinant 
appearing in the denominator of a tensor coefficient is compensated by
a delicate numerical cancellation between scalar integrals in the
numerator. The situation is not very problematic in $2\to2$ reactions,
where the boundary is reached in forward and backward scattering, \ie
in isolated points in a single phase-space variable (scattering angle
$\theta=0,\pi$). In the three-particle phase space the situation is more 
involved, and the numerical problems caused by vanishing Gram
determinants of four momenta turn out to be serious.
Fortunately the appearance of these Gram determinants can be avoided
completely if the four-dimensionality of space time is exploited.
In the following we describe such a procedure.

Other methods for calculating tensor 5-point integrals via reduction to
simpler integrals have been described in \citere{Davydychev:1991va};
however, in these approaches either Gram determinants were not avoided 
completely, or the set of standard integrals had to be extended.
Various treatments of scalar 5-point integrals have been presented in
\citeres{Davydychev:1991va,Me65,vanNeerven:1983vr,Denner:kt,Suzuki:2002js}.
Methods for the numerical integration of scalar 5- and 6-point integrals 
\cite{Binoth:2002xh} and of general one-loop integrals with up 
to six external legs \cite{Ferroglia:2002mz} with special
treatments of singularities have been proposed recently.

It is known since a long time that the scalar 5-point integral can be
reduced to scalar 4-point integrals by using the linear dependence of
the integration momentum on the four external momenta in four 
space--time dimensions 
\cite{Me65,vanNeerven:1983vr}. 
In \citere{Denner:kt} a generalization
of this reduction formula to tensor 5-point integrals has been
proposed, but some extra terms were missing in the derivation.
We supplement the derivation of \citere{Denner:kt} by these missing
terms and work out explicit formulas for the tensor coefficients
up to rank~4. 
The relevant formulas for the reduction of the coefficients of the
1-, 2-, 3-, and 4-point functions to standard scalar integrals are also
presented. Apart from these standard functions no integrals are needed.
In summary, 
the results of this paper comprise all one-loop tensor integrals
that occur in $1\to2$, $2\to2$, and $2\to3$ particle reactions with
up to four external gauge bosons 
(and an arbitrary number of spin-$\frac{1}{2}$ fermions and scalars)
in renormalizable gauge theories.
Apart from the six-point functions, the presented set
covers also all one-loop tensor integrals appearing in processes
with six external particles including up to 
two external bosons, 
i.e., in particular, the processes
$2\Pf\to4\Pf$ and $2\Pf\to2\Pf+2V$.
Only scalar 1-, 2-, 3-, and 4-point integrals have to be calculated in
addition; to this end, methods and explicit results can be found
in the literature
\cite{Davydychev:1991va,Denner:kt,'tHooft:1978xw,Beenakker:1990jr}.

In its present formulation, the described approach applies to
tensor 5-point integrals in four space--time dimensions, \ie
possible infrared (IR)
singularities of soft or collinear origin are assumed to be
regularized by off-shell or mass regulators.%
\footnote{In 5-point functions, ultraviolet (UV) divergences appear only for
tensor rank $\ge6$, which is not considered in this paper.}
This procedure is widely used in the calculation of electroweak
corrections. However, the present formalism can also be used in
dimensional regularization, which is usually adopted in QCD for
the treatment of IR singularities. The translation of 
mass regulators into dimensional regularization is generally
described in \citere{Catani:2001ef} for complete QCD and SUSY-QCD 
amplitudes. A possible strategy for the corresponding treatment
of individual integrals can be found in \citere{Beenakker:2001rj}.

The method proposed in
this paper has already been tested in actual calculations: 
it has been used in the calculation of 
the QCD corrections to the process 
$gg/q\bar q\to\Pt\bar\Pt\PH$
\cite{Beenakker:2001rj}
and the electroweak corrections to $\Pep\Pem\to\nu_l\bar\nu_l\PH$
\cite{eennH}.
The results are in numerical agreement with an evaluation that is based on
the usual Passarino--Veltman reduction, but a drastic
improvement in the numerical stability of the 5-point function was
observed. 

The paper is organized as follows.
In \refse{se:derivation} we derive the general relation between
tensor 5-point integrals and the related 4-point functions. The
explicit results for the tensor coefficients of the 5-point 
integrals up to rank~4 are presented in \refse{se:E_results}.
In \refse{se:checks} we describe some consistency checks and
applications of the presented results; details on the generalizations
of the method to dimensional regularization can also be found there.
Section~\ref{se:concl} contains our conclusion.
In \refapp{app:UVterms} we describe the derivation of extra terms
that arise from UV divergences in intermediate expressions.
Appendices~\ref{app:1234reduction} and \ref{sectendiv} comprise
useful results for 1-, 2-, 3-, and 4-point functions.

\section{Derivation of the reduction formula for tensor 5-point integrals}
\label{se:derivation}

In the reduction of tensor 5-point integrals
to 4-point integrals we closely follow the strategy and notation
of \citeres{Denner:kt,Denner:1998ia}.
The one-loop 4- and 5-point functions are defined as
\beqar \label{CDEint}
\lefteqn{D_{\{0,\mu,\mu\nu,\mu\nu\rho,\mu\nu\rho\si,\mu\nu\rho\si\tau\}}
(p_1,p_2,p_3,m_0,m_1,m_2,m_3)}\qquad\nl
&=&
\frac{(2\pi\mu)^{(4-D)}}{\ri\pi^{2}}\int\!\rd^{D}q\,
\frac{\{1,q_\mu,q_\mu q_\nu,q_\mu
  q_\nu q_\rho, q_\mu q_\nu q_\rho q_\si, q_\mu q_\nu q_\rho q_\si q_\tau\}}{N_0 N_1 N_2 N_3},
\nl
\lefteqn{E_{\{0,\mu,\mu\nu,\mu\nu\rho,\mu\nu\rho\si\}}(p_1,p_2,p_3,p_4,m_{0},m_1,m_2,m_3,m_4)}
\qquad \nl
&=&
\frac{1}{\ri\pi^{2}}\int\!\rd^{4}
q\,\frac{\{1,q_\mu,q_\mu q_\nu,q_\mu
  q_\nu q_\rho, q_\mu q_\nu q_\rho q_\si\}}{N_0 N_1 N_2 N_3 N_4},
\eeqar
with the denominator factors
\beq \label{D0Di}
N_{i}= (q+p_{i})^{2}-m_{i}^{2}+\ri\epsilon, \qquad i=0,\ldots,4 ,
\qquad p_0=0,
\eeq
where $\ri\epsilon$ $(\epsilon>0)$ denotes an infinitesimal imaginary
part. Apart from $D_{\mu\nu\rho\si}$ and $D_{\mu\nu\rho\si\tau}$, all
above tensor integrals are UV finite and can be evaluated in four
dimensions ($D=4)$.

The reduction of the 5-point function to 4-point functions is
based on the fact that in four dimensions the integration momentum $q$
depends linearly on the four external momenta $p_i$ \cite{Me65}.
This gives rise to the identity
\beq\label{detid}
0=\left\vert
\barr{cccc}
2q^{2}  & 2qp_{1}    & \ldots & \;2qp_{4} \\
2p_{1}q & 2p_{1}p_{1} & \ldots & \;2p_{1}p_{4} \\
\vdots    & \vdots     & \ddots     &\;\vdots     \\
2p_{4}q & 2p_{4}p_{1} & \ldots &\; 2p_{4}p_{4}
\earr
\right\vert =
\left\vert\barr{cccc}
2N_{0}+Y_{00}  & 2qp_{1}    & \ldots & \;2qp_{4} \\
N_{1}-N_{0}+Y_{10}-Y_{00} & 2p_{1}p_{1} & \ldots & \;2p_{1}p_{4} \\
\vdots    & \vdots     & \ddots     &\;\vdots     \\
N_{4}-N_{0}+Y_{40}-Y_{00} & 2p_{4}p_{1} & \ldots &\; 2p_{4}p_{4}
\earr\right\vert
\eeq
with
\beq\label{defY}
Y_{ij} = m_i^2 + m_j^2 - (p_i-p_j)^2, \qquad i,j=0,\ldots,4.
\eeq
Equation \refeq{detid} implies
\beq\label{eq:Ered0}
0 = 
\frac{1}{\ri\pi^{2}}\int\! \rd^{4}q\, \frac{q_{\mu_1}\ldots
  q_{\mu_P}}{N_{0}N_{1}\cdots N_{4}}
\frac{-\La^2}{q^2-\La^2}
\left\vert
\barr{cccc}
2N_{0}+Y_{00}  & 2qp_{1}    & \ldots & \;2qp_{4} \\
N_{1}-N_{0}+Y_{10}-Y_{00} & 2p_{1}p_{1} & \ldots &\; 2p_{1}p_{4} \\
\vdots    & \vdots     & \ddots     &\;\vdots     \\
N_{4}-N_{0}+Y_{40}-Y_{00} & 2p_{4}p_{1} & \ldots &\; 2p_{4}p_{4}
\earr
\right\vert,
\eeq
where $P$ denotes the number of integration momenta in the numerator.
UV divergences that occur in intermediate steps by expanding the
determinant are regularized temporarily with a cutoff $\La\to\infty$,
in order to be able to exploit the four-dimensionality of space--time.
This approach is valid for all considered 5-point integrals, \ie for
$P\le4$.  Expanding the determinant along the first column, we obtain
\beqar \label{eq:Ered1}
0&=&\Bigl[2D^{(\La)}_{\mu_1\ldots\mu_P}(0)+Y_{00}E^{(\La)}_{\mu_1\ldots\mu_P}\Bigr]
\left\vert\barr{ccc}
 2p_{1}p_{1}    & \ldots & \;2p_{1}p_{4} \\
 \vdots     & \ddots     &\;\vdots     \\
 2p_{4}p_{1} & \ldots &\; 2p_{4}p_{4}
\earr\right\vert 
\nl&&{}
+\disp\sum_{i=1}^{4}(-1)^{i}\Bigl\{D^{(\La)}_{\al\mu_1\ldots\mu_P}(i)
-
D^{(\La)}_{\al\mu_1\ldots\mu_P}(0)
+(Y_{i0}-Y_{00})E^{(\La)}_{\al\mu_1\ldots\mu_P}
\Bigr\}
\nl&&{}\qquad\times
\left\vert\barr{ccc}
 2p_1^\al    & \ldots & \;2p_4^\al \\
 2p_1p_1 & \ldots &\; 2p_1p_4\\
 \vdots     & \ddots     &\;\vdots     \\
 2p_{i-1}p_1 & \ldots &\; 2p_{i-1}p_4\\
 2p_{i+1}p_1 & \ldots &\; 2p_{i+1}p_4\\
 \vdots     & \ddots     &\;\vdots     \\
 2p_4p_1 & \ldots &\; 2p_4p_4
\earr\right\vert,
\eeqar
where $D^{(\La)}_{\mu_1\ldots\mu_P}(i)$ denotes the 4-point function
that is obtained from the 5-point function
$E^{(\La)}_{\mu_1\ldots\mu_P}$ by omitting the $i$th propagator
$N_i^{-1}$. The superscript $(\La)$ indicates the regularization as
introduced in \refeq{eq:Ered0}. Since all appearing
5-point functions are UV finite ($\mathrm{rank}\le5$), 
we can directly perform the limit $\La\to\infty$ there and omit the 
superscript $(\La)$ for all $E$ functions. The same
can be done for the UV-finite 4-point functions. The UV-divergent
4-point functions are for asymptotically large $\La$ 
split as follows,
\beq
D^{(\La)}_{\mu_1\ldots\mu_P}\asymLa D^{\fin}_{\mu_1\ldots\mu_P} +
\De_{\mu_1\ldots\mu_P},
\label{eq:DLambda}
\eeq
where $D^{\fin}_{\mu_1\ldots\mu_P}$ is the UV-finite part of the usual
4-point function $D_{\mu_1\ldots\mu_P}$
in $D$ dimensions defined in \refeq{CDEint}.  The UV
divergence is subtracted from $D_{\mu_1\ldots\mu_P}$
as in the $\overline{\mathrm{MS}}$ scheme.%
\footnote{The final results for the 5-point functions
  $E_{\mu_1\ldots\mu_P}$ do not depend on the details of this
  subtraction, because these results are finite. We perform the
  subtraction in order to avoid confusion in products involving metric
  tensors. Since the subtraction guarantees that all appearing
  coefficients are finite, the question of whether we take the metric
  tensor in 4 or $D$ dimensions concerns only irrelevant terms of
  $\O(D-4)$.}  Whenever the functions $D_{\mu_1\ldots\mu_P}$ (or their
coefficients) are finite, $D_{\mu_1\ldots\mu_P}$ and
$D^{\fin}_{\mu_1\ldots\mu_P}$ are identical, and we simply write
$D_{\mu_1\ldots\mu_P}$.  The terms $\De_{\mu_1\ldots\mu_P}$ contain
the dependence on $\La$, but are also finite for finite $\Lambda$.
Inserting \refeq{eq:DLambda} into \refeq{eq:Ered1} and taking $\La$
asymptotically large, results in
\beqar \label{eq:Ered1a}
0&=&\Bigl[2D^{\fin}_{\mu_1\ldots\mu_P}(0)+2\De_{\mu_1\ldots\mu_P}(0)
+Y_{00}E_{\mu_1\ldots\mu_P}\Bigr]
\left\vert\barr{ccc}
 2p_{1}p_{1}    & \ldots & \;2p_{1}p_{4} \\
 \vdots     & \ddots     &\;\vdots     \\
 2p_{4}p_{1} & \ldots &\; 2p_{4}p_{4}
\earr\right\vert \nl
&&+\disp\sum_{i=1}^{4}(-1)^{i}\Bigl\{D^{\fin}_{\al\mu_1\ldots\mu_P}(i) 
-\Bigl[D^{\fin}_{\al\mu_1\ldots\mu_P}(0)
+p_{1\al}D^{\fin}_{\mu_1\ldots\mu_P}(0)\Bigr]
+p_{1\al}D^{\fin}_{\mu_1\ldots\mu_P}(0)
\nl&&{}
+\De_{\al\mu_1\ldots\mu_P}(i)-\De_{\al\mu_1\ldots\mu_P}(0)
+(Y_{i0}-Y_{00})E_{\al\mu_1\ldots\mu_P}
\Bigr\}
\left\vert\barr{ccc}
 2p_1^\al    & \ldots & \;2p_4^\al \\
 2p_1p_1 & \ldots &\; 2p_1p_4\\
 \vdots     & \ddots     &\;\vdots     \\
 2p_{i-1}p_1 & \ldots &\; 2p_{i-1}p_4\\
 2p_{i+1}p_1 & \ldots &\; 2p_{i+1}p_4\\
 \vdots     & \ddots     &\;\vdots     \\
 2p_4p_1 & \ldots &\; 2p_4p_4
\earr\right\vert.
\eeqar
The terms involving $p_{1\al}D^{\fin}_{\mu_1\ldots\mu_P}(0)$ have been
added for later convenience. The extra terms involving
$\De_{\mu_1\ldots\mu_P}$ are absent for $P\le2$ where no UV-singular
integrals appear; they drop out in the final result for $P=3$, but
contribute for $P=4$.

The contributions to 4-point functions that involve a momentum
$p_{j\al}$ will be simplified in the following.
To this end, we introduce the Lorentz-covariant decompositions 
\beqar\label{decomp}
D^{\fin}_{\al\mu_1\ldots\mu_P}(i) &=& [D^{\fin}_{\al\mu_1\ldots\mu_P}(i)]^{(p)}
+ [D^{\fin}_{\al\mu_1\ldots\mu_P}(i)]^{(g)}, \qquad i=0,\ldots,4,\nl{}
 [D^{\fin}_{\al\mu_1\ldots\mu_P}(i)]^{(p)}&=&\sum_{j=1\atop j\ne i}^4 
p_{j\al}X_{j,\mu_1\ldots\mu_P}(i),\nl{}
 [D^{\fin}_{\al\mu_1\ldots\mu_P}(i)]^{(g)}&=&\sum_{j=1}^P 
g_{\al\mu_j}Y_{j,\mu_1\ldots\mu_{j-1}\mu_{j+1}\ldots\mu_P}(i),\nl{}
\lefteqn{[D^{\fin}_{\al\mu_1\ldots\mu_P}(0)
+p_{1\al} D^{\fin}_{\mu_1\ldots\mu_P}(0)]^{(p)}=
\sum_{j=2}^4 
(p_{j}-p_{1})_\al Z_{\mu_1\ldots\mu_P}.}
\phantom{[D_{\al\mu_1\ldots\mu_P}(i)]^{(p)}}
\eeqar
The operation ``$(g)$'' isolates all tensor structures in
which the first Lorentz index appears at a metric tensor; the
remaining part of the tensor furnishes the ``$(p)$'' contribution.
The last decomposition in \refeq{decomp} becomes obvious after
performing a shift $q\to q-p_1$ in the integral. From \refeq{decomp}
it follows immediately that the terms in \refeq{eq:Ered1a} that
involve $[D^{\fin}_{\al\mu_1\ldots\mu_P}(i)]^{(p)}$ 
drop out when
multiplied with the determinants, because the resulting determinants
vanish. Similarly, the contribution proportional to
$[D^{\fin}_{\al\mu_1\ldots\mu_P}(0)
+p_{1\al}D^{\fin}_{\mu_1\ldots\mu_P}(0)]^{(p)}$ 
vanishes after summation over $i$.  Finally, the term
$p_{1\al}D^{\fin}_{\mu_1\ldots\mu_P}(0)$ contributes only for $i=1$,
where it can be combined with the first term in (\ref{eq:Ered1a}).
Rewriting the resulting equation using determinants and reinserting
the explicit
form of the tensor integrals leads to%
\footnote{In \citere{Denner:kt} the $V$ and $U$ terms are missing.}
\beq \label{eq:Ered2}
\left[\frac{(2\pi\mu)^{4-D}}{\ri\pi^{2}}\int\!\rd^{D}q\,
\frac{q_{\mu_1}\ldots q_{\mu_P}}{N_{0}N_{1}\cdots N_{4}}
\left\vert
\barr{cccc}
N_{0}+Y_{00}  & 2qp_{1}    & \ldots & \;2qp_{4} \\
Y_{10}-Y_{00} & 2p_{1}p_1 & \ldots & \;2p_{1}p_{4} \\
\vdots    & \vdots     & \ddots     &\;\vdots     \\
Y_{40}-Y_{00} & 2p_{4}p_{1} & \ldots &\; 2p_{4}p_4
\earr
\right\vert\;\right]^{\fin}
=
\extradet_{\mu_1\ldots\mu_P} + \UVdet_{\mu_1\ldots\mu_P}.
\eeq
Here we introduced
\beq
\extradet_{\mu_1\ldots\mu_P}=
-\left\vert
\barr{cccc}
           0                & 2p_{1}^\al& \ldots & \;2p_{4}^\al \\
\D_{\al\mu_1\ldots\mu_P}(1) & 2p_{1}p_1 & \ldots & \;2p_{1}p_{4} \\
\vdots    & \vdots     & \ddots     &\;\vdots     \\
\D_{\al\mu_1\ldots\mu_P}(4) & 2p_{4}p_1 & \ldots & \;2p_{4}p_{4} 
\earr
\right\vert
=\sum_{i,j=1}^4 (-1)^{i+j} \, \det(\Zhat^{(4)}_{ij})\,
2p^\al_j\D_{\al\mu_1\ldots\mu_P}(i)
\eeq
with
\beq
\D_{\al\mu_1\ldots\mu_P}(i) =
[D^{\fin}_{\al\mu_1\ldots\mu_P}(i)-D^{\fin}_{\al\mu_1\ldots\mu_P}(0)]^{(g)}, 
\qquad i=1,\dots,4,
\eeq
which collects all terms involving $[D]^{(g)}$.
The extra term
$\extradet$ is absent for the reduction of the scalar
5-point integral since $D_\al=[D_\al]^{(p)}$.  
The three-dimensional
matrices $\Zhat^{(4)}_{ij}$ result from the four-dimensional Gram
matrix
\beq
Z^{(4)}=\left(
\barr{ccc}
 2p_{1}p_1 & \ldots & \;2p_{1}p_{4} \\
\vdots     & \ddots     &\;\vdots   \\
 2p_{4}p_1 & \ldots & \;2p_{4}p_{4} 
\earr
\right)
\eeq
by discarding the $i$th row and $j$th column.
The term $\UVdet_{\mu_1\ldots\mu_P}$ resulting from the UV divergences reads
\beq\label{eq:UVdetdef}
\UVdet_{\mu_1\ldots\mu_P}= -2\De_{\mu_1\ldots\mu_P}(0)
\det(Z^{(4)})
+ \sum_{i,j=1}^4 (-1)^{i+j} \, \det(\Zhat^{(4)}_{ij})\,
2p^\al_j[\De_{\al\mu_1\ldots\mu_P}(i)-\De_{\al\mu_1\ldots\mu_P}(0)].
\eeq
The actual evaluation of $\UVdet_{\mu_1\ldots\mu_P}$ is described in
\refapp{app:UVterms} and yields
\beq\label{eq:UVdet}
\UVdet_{\mu_1\ldots\mu_P}=\left\{\barr{ll} 0 & \quad \mathrm{for\ } P\le 3,\nl
               \displaystyle -\textfrac{1}{48}
 (g_{\mu_1\mu_2}g_{\mu_3\mu_4}+g_{\mu_1\mu_3}g_{\mu_4\mu_2}
 +g_{\mu_1\mu_4}g_{\mu_2\mu_3})\det(Z^{(4)}) &
 \quad \mathrm{for\ } P=4. \earr
 \right. 
\eeq
We did not investigate these terms for $P\ge5$.

Using
\beq
2p_{i}p_{j}= Y_{ij}-Y_{i0}-Y_{0j}+Y_{00}, \qquad
2qp_{j}= N_{j}-N_{0}+Y_{0j}-Y_{00},
\eeq
we can transform the determinant on the left-hand side of
\refeq{eq:Ered2} by adding the first column to each of the other
columns, and then enlarging the determinant by one column and one row,
resulting in
\beq\label{eq:Ered3}
\left\vert\barr{cccc}
1 & Y_{00}    & \ldots & \;Y_{04} \\
0 & \quad D^{\fin}_{\mu_1\ldots\mu_P}(0)+Y_{00}E_{\mu_1\ldots\mu_P} \quad & 
\ldots & \quad D^{\fin}_{\mu_1\ldots\mu_P}(4)+Y_{04}E_{\mu_1\ldots\mu_P} \\
0 & Y_{10}-Y_{00} & \ldots &\; Y_{14}-Y_{04}  \\
\vdots    & \vdots     & \ddots     &\;\vdots     \\
0 & Y_{40}-Y_{00} & \ldots &\; Y_{44}-Y_{04}
\earr\right\vert
=\extradet_{\mu_1\ldots\mu_P}+\UVdet_{\mu_1\ldots\mu_P}.
\eeq
Equation \refeq{eq:Ered3} is equivalent to
\beqar \label{eq:Ered}
\lefteqn{\left\vert \barr{cccccc}
E_{\mu_1\ldots\mu_P}
&\:-D^{\fin}_{\mu_1\ldots\mu_P}(0)&\:-D^{\fin}_{\mu_1\ldots\mu_P}(1)&
\:-D^{\fin}_{\mu_1\ldots\mu_P}(2)&\:-D^{\fin}_{\mu_1\ldots\mu_P}(3)&
\:-D^{\fin}_{\mu_1\ldots\mu_P}(4)\\
  1   &  Y_{00}   &  Y_{01}   &  Y_{02}   &  Y_{03}   &  Y_{04}   \\
  1   &  Y_{10}   &  Y_{11}   &  Y_{12}   &  Y_{13}   &  Y_{14}   \\
  1   &  Y_{20}   &  Y_{21}   &  Y_{22}   &  Y_{23}   &  Y_{24}   \\
  1   &  Y_{30}   &  Y_{31}   &  Y_{32}   &  Y_{33}   &  Y_{34}   \\
  1   &  Y_{40}   &  Y_{41}   &  Y_{42}   &  Y_{43}   &  Y_{44}
\earr \right\vert
}\quad &&\phantom{-E_{\mu_1\ldots\mu_P}
\:\:D_{\mu_1\ldots\mu_P}(0)\:\:D_{\mu_1\ldots\mu_P}(1)\:
\:D_{\mu_1\ldots\mu_P}(2)\:\:D_{\mu_1\ldots\mu_P}(3)\:
\:D_{\mu_1\ldots\mu_P}(4)} \quad\nl[2ex]
&=&\extradet_{\mu_1\ldots\mu_P}+\UVdet_{\mu_1\ldots\mu_P},
\eeqar
which expresses the tensor 5-point function $E_{\mu_1\ldots\mu_P}$ in terms of
five tensor 4-point functions
\beqar\label{eq:Eredf1}
 E_{\mu_1\ldots\mu_P} &=& - \sum_{i=0}^4
 \,\frac{\det(Y_i)}{\det(Y)} \,D^{\fin}_{\mu_1\ldots\mu_P}(i)
+\sum_{i,j=1}^4
(-1)^{i+j} \, \frac{\det(\Zhat^{(4)}_{ij})}{\det(Y)}\,2p^\al_j\D_{\al\mu_1\ldots\mu_P}(i)
\nl
&& {}+\frac{1}{\det(Y)}\UVdet_{\mu_1\ldots\mu_P},
\eeqar
where $Y=(Y_{ij})$, $i=0,\ldots4$, and $Y_i$ is obtained from the
5-dimensional Cayley matrix $Y$ by replacing all entries in the $i$th
column with~1.

With the shorthand notation
\beq
\detY{i}=\frac{\det(Y_i)}{\det(Y)}, \quad 
i=0,\ldots,4, \qquad
\detZh{ij} = (-1)^{i+j} \, 
\frac{\det(\Zhat_{ij}^{(4)})}{\det(Y)}=\detZh{ji}, \quad i,j=1,\ldots,4,
\eeq
this reads
\beq\label{eq:Eredf}
 E_{\mu_1\ldots\mu_P} = - \sum_{i=0}^4
 \detY{i}D^{\fin}_{\mu_1\ldots\mu_P}(i)
+2\sum_{i,j=1}^4 \detZh{ij}\,p^\al_j\,\D_{\al\mu_1\ldots\mu_P}(i)
+\frac{1}{\det(Y)}\UVdet_{\mu_1\ldots\mu_P}
\eeq
with $\UVdet$ given in \refeq{eq:UVdet} for $P\le4$.

\section{Explicit formulas for 5-point tensor coefficients}
\label{se:E_results}

Here we further exploit \refeq{eq:Eredf} to derive explicit formulas
for the coefficients of tensor 5-point integrals appearing in
convenient decompositions into Lorentz covariants. 
In order to be able to write down the tensor decompositions in a
concise way we introduce the notation
\beq\label{eq:abbr3}
T^{[\mu_1\ldots\mu_P]} = T^{\mu_1\ldots\mu_P}+
T^{\mu_2\ldots\mu_P\mu_1}+ \ldots +T^{\mu_P\mu_1\ldots\mu_{P-1}},
\eeq
\ie Lorentz indices within square brackets 
represent a sum over all tensors with
cyclic permutations of these indices. 
For example, we have
\beq
g_{[\mu\nu}g_{\rho]\si}=g_{\mu\nu}g_{\rho\si}+g_{\nu\rho}g_{\mu\si}
+g_{\rho\mu}g_{\nu\si}.
\eeq
This notation can be iterated, \eg,
\beq
T^{[\mu[\nu\rho]]} =
T^{[\mu\nu\rho]}+ T^{[\mu\rho\nu]}
=T^{\mu\nu\rho}+T^{\nu\rho\mu}+T^{\rho\mu\nu}
+ T^{\mu\rho\nu}+ T^{\rho\nu\mu}+ T^{\nu\mu\rho}.
\eeq

After cancelling the denominator $N_0$ the resulting tensor integrals
are not in the standard form  but can be expressed in terms of standard
integrals by shifting the integration momentum. We choose to perform
the shift $q\to q-p_1$, so that the following 4-point integrals appear:
\beqar \label{Dshifted}
\Dtilde_{\{0,\mu,\mu\nu,\mu\nu\rho,\mu\nu\rho\si,\mu\nu\rho\si\tau\}}(0)
&=&
\frac{(2\pi\mu)^{(4-D)}}{\ri\pi^{2}}\int\!\rd^{D}q\,
\frac{\{1,q_\mu,q_\mu q_\nu,q_\mu q_\nu q_\rho, q_\mu q_\nu q_\rho q_\si,
   q_\mu q_\nu q_\rho q_\si q_\tau\}}{\Ntilde_1 \Ntilde_2
  \Ntilde_3 \Ntilde_4},
\nl
\Ntilde_{i}&=& (q+p_{i}-p_{1})^{2}-m_{i}^{2}+\ri\epsilon, \qquad i=1,\ldots,4 .
\eeqar
Note that 
the scalar integral 
$D_{0}$, 
and the tensor coefficients 
$D_{00}$ and $D_{0000}$ are invariant under this
shift. Therefore, we can omit the tilde on these functions. 
In the decomposition of
$D_{\{\mu,\mu\nu,\mu\nu\rho,\mu\nu\rho\si,\mu\nu\rho\si\tau\}}(i)$
with $i=1,\dots,4$ shifted indices appear which we denote as
\beq\label{eq:defji}
 j_i=\left\{\barr{lll} j & \mathrm{\ for\ }&  i>j,\\
                     j-1 & \mathrm{\ for\ }& i < j.\earr \right.
\eeq
We also use the notation $\debar_{ij} = 1-\de_{ij}$, \ie
$\sum_i \debar_{ij}(\dots) = \sum_{i\ne j}(\dots)$.

\paragraph{Scalar integral}

For the scalar 5-point function \refeq{eq:Eredf} reads
\beq\label{E0red}
 E_{0} = - \sum_{i=0}^4\detY{i}D_{0}(i),
\eeq
as was also obtained in 
\citeres{Me65,vanNeerven:1983vr,Denner:kt,Denner:1998ia}.

\paragraph{Vector integral}

For the vector 5-point function and the relevant 4-point functions we
have the following covariant decompositions
\beqar
E^\mu&=&\sum_{j=1}^4 p_j^{\mu}E_j,\nl
D^\mu(i)&=&\sum_{j=1\atop j\ne i}^4 p_j^{\mu}D_{j_i}(i),\qquad i=1,\dots,4,
\nl
D^\mu(0)&=&\Dtilde^\mu(0)-p_1^{\mu}D_0(0)=
\sum_{j=2}^4 (p_{j}-p_{1})^\mu\Dtilde_{j-1}(0)-p_1^{\mu}D_0(0),\nl
\D^{\al\mu}(i)&=&g^{\al\mu}[D_{00}(i)-D_{00}(0)].
\eeqar
Inserting these into \refeq{eq:Eredf} for the vector integral, we find for
the components
\beqar
E_{j}&=& - \sum_{i=1}^4\detY{i} D_{j_i}(i)\debar_{ji}
-\detY{0}\Dcomb_{j}(0)
+2\sum_{i=1}^4 \detZh{ji} [D_{00}(i)-D_{00}(0)],
\qquad j=1,\dots,4,
\eeqar
with
\beqar
\Dcomb_{1}(0) &=& -\sum_{j=1}^3\Dtilde_{j}(0)-D_{0}(0),\qquad
\Dcomb_{j}(0) = \Dtilde_{j-1}(0), \qquad j=2,3,4.
\eeqar

\paragraph{Tensor integral of rank 2}

For the case of the second rank tensor we introduce the
covariant decompositions
\beqar
E^{\mu\nu}&=&\sum_{j,k=1}^4 p_j^{\mu}p_k^{\nu}E_{jk}+g^{\mu\nu}E_{00},\nls
D^{\mu\nu}(i)&=&\sum_{j,k=1\atop j,k\ne i}^4 p_j^{\mu}p_k^{\nu}D_{j_i
  k_i}(i) + g^{\mu\nu}D_{00}(i),\qquad i=1,\dots,4,\nls
D^{\mu\nu}(0)&=&\Dtilde^{\mu\nu}(0)-p_1^{[\mu}\Dtilde^{\nu]}(0)
+p_1^{\mu}p_1^{\nu}D_{0}(0)\nl
&=&
\sum_{j,k=2}^4 (p_{j}-p_{1})^\mu(p_{k}-p_{1})^\nu\Dtilde_{j-1,k-1}(0)
+ g^{\mu\nu}D_{00}(0)\nl
&&{}-\sum_{j=2}^4 p_1^{[\mu}(p_{j}-p_{1})^{\nu]}\Dtilde_{j-1}(0)
+p_1^{\mu}p_1^{\nu}D_{0}(0),\nls
\D^{\al\mu\nu}(i)&=&
\Bigl[D^{\al\mu\nu}(i) -D^{\al\mu\nu}(0) \Bigr]^{(g)}\nl
&=& \Bigl[D^{\al\mu\nu}(i)
-\Dtilde^{\al\mu\nu}(0) + \Dtilde^{\al[\nu}(0) p_1^{\mu]} \Bigr]^{(g)}\nl
&=& \sum_{j=1\atop j\ne
  i}^4g^{\al[\mu}p_j^{\nu]}D_{00j_i}(i)
-\sum_{j=2}^4g^{\al[\mu}(p_j-p_1)^{\nu]}\Dtilde_{00,j-1}(0)
+p_1^{[\mu}g^{\nu]\al}D_{00}(0).
\eeqar
The $E_{00}$ term in the decomposition of $E^{\mu\nu}$ is redundant
in the sense that the corresponding covariant, $g^{\mu\nu}$,
can be expressed by the four linearly independent vectors $p_j$
up to terms of $\O(D-4)$,
\beq\label{gred}
g^{\mu\nu}= \sum_{j,k=1}^4 2p_j^{\mu}p_k^{\nu} (Z^{(4)})^{-1}_{jk}
+\O(D-4).
\eeq
This means that the set of covariants in the decomposition of
$E^{\mu\nu}$ is overcomplete, and the coefficient $E_{00}$ can be
defined by convenience. We use this freedom to avoid 
$\det(Z^{(4)})$ in the denominators.

With the above decompositions we find from \refeq{eq:Eredf} 
\beqar
E_{00}&=& - \sum_{i=1}^4\detY{i} D_{00}(i)
-\detY{0}D_{00}(0),\nnls
E_{jk}&=&
2\left\{\sum_{i=1}^4 \detZh{ji} 
[D_{00k_i}(i)\debar_{ki}-\Dcomb_{00k}(0)] + (j\leftrightarrow
k)\right\}
\nl&&{}
 - \sum_{i=1}^4\detY{i} D_{j_ik_i}(i)\debar_{ji}\debar_{ki}
-\detY{0}\Dcomb_{jk}(0),\qquad j,k=1,\dots,4,
\eeqar
with
\beqar
\Dcomb_{11}(0) &=& \sum_{j,k=1}^3\Dtilde_{jk}(0)
+2\sum_{j=1}^3\Dtilde_{j}(0)+D_{0}(0),\nl
\Dcomb_{j1}(0) &=& -\sum_{k=1}^3\Dtilde_{j-1,k}(0)-\Dtilde_{j-1}(0),
\nl
\Dcomb_{jk}(0) &=& \Dtilde_{j-1,k-1}(0), 
\nnls
\Dcomb_{001}(0) &=& -\sum_{j=1}^3\Dtilde_{00j}(0) - D_{00}(0),\nl
\Dcomb_{00j}(0) &=& \Dtilde_{00,j-1}(0), 
\qquad j,k=2,3,4.
\eeqar

\paragraph{Tensor integral of rank 3}

The covariant decompositions of the integrals appearing in the
reduction of the 3rd rank tensor read
\beqar
E^{\mu\nu\rho}&=&\sum_{j,k,l=1}^4 p_j^{\mu}p_k^{\nu}p_l^{\rho}E_{jkl}
+\sum_{j=1}^4g^{[\mu\nu} p_j^{\rho]}E_{00j},\nls
D^{\mu\nu\rho}(i)&=&\sum_{j,k,l=1\atop j,k,l\ne i}^4
p_j^{\mu}p_k^{\nu}p_l^{\rho}D_{j_i k_i l_i}(i) 
+ \sum_{j=1\atop j\ne i}^4 g^{[\mu\nu} p_j^{\rho]}\Dtilde_{00j_i}(i),
\qquad i=1,\dots,4,\nls
D^{\mu\nu\rho}(0)&=&\Dtilde^{\mu\nu\rho}(0)-p_1^{[\mu}\Dtilde^{\nu\rho]}(0)
+p_1^{[\mu}p_1^{\nu}\Dtilde^{\rho]}(0)-p_1^{\mu}p_1^{\nu}p_1^{\rho}D_{0}(0)
\nl
&=&
\sum_{j,k,l=2}^4
(p_{j}-p_{1})^\mu(p_{k}-p_{1})^\nu(p_{l}-p_{1})^\rho\Dtilde_{j-1,k-1,l-1}(0)
\nl&&{}
+ \sum_{j=2}^4 g^{[\mu\nu} (p_{j}-p_{1})^{\rho]}\Dtilde_{00,j-1}(0) \nl
&&{}-p_1^{[\mu}\sum_{j,k=2}^4(p_j-p_1)^\nu(p_{k}-p_{1})^{\rho]}
\Dtilde_{j-1,k-1}(0)-p_1^{[\mu} g^{\nu\rho]}D_{00}(0)\nl
&&{}+p_1^{[\mu}p_1^{\nu}\sum_{j=2}^4(p_{j}-p_{1})^{\rho]}\Dtilde_{j-1}(0)
-p_1^{\mu}p_1^{\nu}p_1^{\rho}D_{0}(0),\nls
\D^{\al\mu\nu\rho}(i)&=&
\Bigl[ D^{\al\mu\nu\rho,\fin}(i)-D^{\al\mu\nu\rho,\fin}(0) \Bigr]^{(g)}\nl
&=& \Bigl[ D^{\al\mu\nu\rho,\fin}(i)-\Dtilde^{\al\mu\nu\rho,\fin}(0)
+\Dtilde^{\al[\nu\rho}(0)p_1^{\mu]}-\Dtilde^{\al[\rho}(0)p_1^{\mu}p_1^{\nu]}
\Bigr]^{(g)}\nl
&=&\sum_{j,k=1\atop j,k\ne i}^4
g^{\al[\mu}p_j^{\nu}p_k^{\rho]}D_{00j_ik_i}(i)
+g^{\al[\mu}g^{\nu\rho]}[D^{\fin}_{0000}(i)-D^{\fin}_{0000}(0)]\nl
&& -\sum_{j,k=2}^4g^{\al[\mu}(p_{j}-p_1)^\nu(p_k-p_1)^{\rho]}
\Dtilde_{00,j-1,k-1}(0)
\nl&&{}
+p_1^{[\mu} \sum_{j=2}^4(p_{j}-p_1)^{[\nu} g^{\rho]]\al}\Dtilde_{00,j-1}(0)
-g^{\al[\mu}p_1^{\nu}p_1^{\rho]}D_{00}(0).
\eeqar
Again the coefficients $E_{00j}$ are redundant and introduced
for convenience. The coefficients of the 3rd rank tensor 5-point
function are then given by
\beqar
E_{00j}&=& - \sum_{i=1}^4\detY{i} D_{00j_i}(i)\debar_{ji}
-\detY{0}\Dcomb_{00j}(0)
+2\sum_{i=1}^4 \detZh{ji} 
[D^{\fin}_{0000}(i)-D^{\fin}_{0000}(0)],
\nnls
E_{jkl}&=& 
\biggl\{2\sum_{i=1}^4 \detZh{ji}
[D_{00k_i l_i}(i)\debar_{ki}\debar_{li}-\Dcomb_{00kl}(0)]
+(j\leftrightarrow k)+(j\leftrightarrow l)\biggr\}
\nl&&{}
- \sum_{i=1}^4
 \detY{i} D_{j_i k_i l_i}(i)\debar_{ji}\debar_{ki}\debar_{li}
-\detY{0}\Dcomb_{jkl}(0)
\qquad j,k,l=1,\dots,4,
\eeqar
with
\beqar
\Dcomb_{111}(0) &=& -\sum_{j,k,l=1}^3\Dtilde_{jkl}(0)
-3\sum_{j,k=1}^3\Dtilde_{jk}(0)-3\sum_{j=1}^3\Dtilde_{j}(0)-D_{0}(0),\nl
\Dcomb_{j11}(0) &=&
\sum_{k,l=1}^3\Dtilde_{j-1,kl}(0)+2\sum_{k=1}^3\Dtilde_{j-1,k}(0)+\Dtilde_{j-1}(0),
\nl
\Dcomb_{jk1}(0) &=&
-\sum_{l=1}^3\Dtilde_{j-1,k-1,l}(0)-\Dtilde_{j-1,k-1}(0), 
\nl
\Dcomb_{jkl}(0) &=& \Dtilde_{j-1,k-1,l-1}(0), 
\nnls
\Dcomb_{0011}(0) &=& \sum_{j,k=1}^3\Dtilde_{00jk}(0)+2\sum_{j=1}^3\Dtilde_{00j}(0)+D_{00}(0),\nl
\Dcomb_{00j1}(0) &=&
-\sum_{k=1}^3\Dtilde_{00,j-1,k}(0)-\Dtilde_{00,j-1}(0), 
\nl
\Dcomb_{00jk}(0) &=& \Dtilde_{00,j-1,k-1}(0), 
\qquad j,k,l=2,3,4.
\eeqar

\paragraph{Tensor integral of rank 4}

For the reduction of the 4th rank tensor we introduce
\beqar
E^{\mu\nu\rho\si}&=&
\sum_{j,k,l,m=1}^4 p_j^{\mu}p_k^{\nu}p_l^{\rho}p_m^{\si}E_{jklm}
+\sum_{j,k=1}^4(g^{[\mu\nu}p_{j}^{\rho]}p_k^\si
+g^{\si[\mu}p_{j}^{\nu}p_k^{\rho]})E_{00jk}
+g^{[\mu\nu}g^{\rho]\si}E_{0000},
\nls
D^{\mu\nu\rho\si,\fin}(i)&=&\sum_{j,k,l,m=1\atop j,k,l,m\ne i}^4
p_j^{\mu}p_k^{\nu}p_l^{\rho}p_m^{\si}D_{j_i k_i l_i m_i}(i) 
+ \sum_{j,k=1\atop j,k\ne i}^4
(g^{[\mu\nu}p_{j}^{\rho]}p_k^\si
+g^{\si[\mu}p_{j}^{\nu}p_k^{\rho]})
D_{00j_ik_i}(i)
\nl&&{}\qquad
+g^{[\mu\nu}g^{\rho]\si}D^{\fin}_{0000}(i),\qquad i=1,\dots,4,\nls
D^{\mu\nu\rho\si,\fin}(0)&=&\Dtilde^{\mu\nu\rho\si,\fin}(0)
-p_1^{[\mu}\Dtilde^{\nu\rho\si]}(0)
+p_1^{[\mu}p_1^{\nu}\Dtilde^{\rho]\si}(0)
+p_1^{\si}p_1^{[\mu}\Dtilde^{\nu\rho]}(0)
-p_1^{[\mu}p_1^{\nu}p_1^{\rho}\Dtilde^{\si]}(0)
\nl&&{}
+p_1^{\mu}p_1^{\nu}p_1^{\rho}p_1^{\si}D_{0}(0)
\nl
&=&
\sum_{j,k,l,m=2}^4
(p_{j}-p_{1})^\mu(p_{k}-p_{1})^\nu(p_{l}-p_{1})^\rho(p_{m}-p_{1})^\si
\Dtilde_{j-1,k-1,l-1,m-1}(0)
\nl&&{}
+ \sum_{j,k=2}^4 [g^{[\mu\nu} (p_{j}-p_{1})^{\rho]}(p_{k}-p_{1})^\si
+g^{\si[\mu} (p_{j}-p_{1})^\nu(p_{k}-p_{1})^{\rho]}]\Dtilde_{00,j-1,k-1}(0)
\nl&&{}
+g^{[\mu\nu}g^{\rho]\si}D^{\fin}_{0000}(0)
-p_1^{[\mu}\sum_{j,k,l=2}^4(p_j-p_1)^\nu(p_k-p_1)^\rho
(p_l-p_1)^{\si]}\Dtilde_{j-1,k-1,l-1}(0)
\nl&&{}
-p_1^{[\mu} \sum_{j=2}^4 g^{[\nu\rho}(p_{j}-p_{1})^{\si]]}\Dtilde_{00,j-1}(0)
\nl&&{} 
+\sum_{j,k=2}^4[p_1^{[\mu}p_1^{\nu}(p_j-p_1)^{\rho]}(p_{k}-p_{1})^\si
+p_1^{\si}p_1^{[\mu}(p_j-p_1)^{\nu}(p_{k}-p_{1})^{\rho]}]
\Dtilde_{j-1,k-1}(0) 
\nl&&{} 
+(p_1^{[\mu}p_1^{\nu}g^{\rho]\si}
+p_1^{\si}p_1^{[\mu}g^{\nu\rho]})D_{00}(0)
\nl&&{} 
-p_1^{[\mu}p_1^{\nu}p_1^{\rho}\sum_{j=2}^4(p_{j}-p_{1})^{\si]}\Dtilde_{j-1}(0)
+p_1^{\mu}p_1^{\nu}p_1^{\rho}p_1^{\si}D_{0}(0),\nl
\D^{\al\mu\nu\rho\si}(i)&=&
\Bigl[ D^{\al\mu\nu\rho\si,\fin}(i)-D^{\al\mu\nu\rho\si,\fin}(0) \Bigr]^{(g)}\nl
&=& \Bigl[ D^{\al\mu\nu\rho\si,\fin}(i)-\Dtilde^{\al\mu\nu\rho\si,\fin}(0)
+\Dtilde^{\al[\nu\rho\si,\fin}(0)p_1^{\mu]}
-\Dtilde^{\al\si[\rho}(0)p_1^{\mu}p_1^{\nu]}
\nl&&{}
-\Dtilde^{\al[\nu\rho}(0)p_1^{\mu]}p_1^{\si}
+\Dtilde^{\al[\si}(0)p_1^{\mu}p_1^{\nu}p_1^{\rho]}
\Bigr] ^{(g)}\nl
&=&\sum_{j,k,l=1\atop j,k,l\ne i}^4
g^{\al[\mu}p_j^{\nu}p_k^{\rho}p_l^{\si]}D_{00j_ik_il_i}(i)
+\sum_{j=1\atop j\ne i}^4
g^{\al[\mu}g^{[\nu\rho}p_j^{\si]]}D^{\fin}_{0000j_i}(i)\nl
&& -\sum_{j,k,l=2}^4 g^{\al[\mu}(p_{j}-p_1)^\nu(p_k-p_1)^\rho (p_l-p_1)^{\si]}
\Dtilde_{00,j-1,k-1,l-1}(0)
\nl&&{}
-\sum_{j=2}^4
g^{\al[\mu}g^{[\nu\rho}(p_{j}-p_1)^{\si]]}\Dtilde^{\fin}_{0000,j-1}(0)
\nl&&{} 
+\sum_{j,k=2}^4
p_1^{[\mu} (p_j-p_1)^{[\nu}(p_{k}-p_1)^{\rho} g^{\si]]\al}\Dtilde_{00,j-1,k-1}(0)
+p_1^{[\mu} g^{[\nu\rho} g^{\si]]\al}D^{\fin}_{0000}(0)
\nl&&{}
-\sum_{j=2}^4
\Bigl[p_1^{[\mu}p_1^{\nu}(p_j-p_1)^{\rho]}g^{\si\al}
+p_1^{[\mu}p_1^{\nu}g^{\rho]\al}(p_j-p_1)^{\si}
\nl&&\qquad{}
+p_1^{\si}p_1^{[\mu}(p_j-p_1)^{[\nu}g^{\rho]]\al}
\Bigr] \Dtilde_{00,j-1}(0)
+p_1^{[\mu}p_1^{\nu}p_1^{\rho}g^{\si]\al} D_{00}(0).
\eeqar

As above, the coefficients $E_{00jk}$ and $E_{0000}$ are redundant and
introduced for convenience. The coefficients of the 4th rank tensor
5-point function are then given by
\beqar
E_{0000}&=& - \sum_{i=1}^4\detY{i} D^{\fin}_{0000}(i)
-\detY{0}D^{\fin}_{0000}(0)-\frac{1}{48}\frac{\det(Z^{(4)})}{\det(Y)},
\nnls
E_{00jk}&=& - \sum_{i=1}^4\detY{i} D_{00j_ik_i}(i)\debar_{ji}\debar_{ki}
-\detY{0}\Dcomb_{00jk}(0)
\nl&&{}
+\left\{2\sum_{i=1}^4 \detZh{ji} 
[D^{\fin}_{0000k_i}(i)\debar_{ki}-\Dcomb^{\fin}_{0000k}(0)]
+(j \leftrightarrow k)\right\},
\nnls
E_{jklm}&=& - \sum_{i=1}^4
 \detY{i} D_{j_i k_i l_i m_i}(i)\debar_{ji}\debar_{ki}\debar_{li}\debar_{mi}
-\detY{0}\Dcomb_{jklm}(0)\nl
&&{}+\biggl\{2\sum_{i=1}^4 \detZh{ji}
[D_{00k_i l_i m_i}(i)\debar_{ki}\debar_{li}\debar_{mi}-\Dcomb_{00klm}(0)]
\nl&&{}\qquad
+(j\leftrightarrow k)+(j\leftrightarrow l)+(j\leftrightarrow m)\biggr\},
\qquad j,k,l,m=1,\dots,4,
\eeqar
with
\beqar
\Dcomb_{1111}(0) &=& \sum_{j,k,l,m=1}^3\Dtilde_{jklm}(0)
+4\sum_{j,k,l=1}^3\Dtilde_{jkl}(0)+6\sum_{j,k=1}^3\Dtilde_{jk}(0)
+4\sum_{j=1}^3\Dtilde_{j}(0)+D_{0}(0),\nl
\Dcomb_{j111}(0) &=& -\sum_{k,l,m=1}^3\Dtilde_{j-1,klm}(0)
-3\sum_{k,l=1}^3\Dtilde_{j-1,kl}(0)-3\sum_{k=1}^3\Dtilde_{j-1,k}(0)
-\Dtilde_{j-1}(0), 
\nl
\Dcomb_{jk11}(0) &=& \sum_{l,m=1}^3\Dtilde_{j-1,k-1,lm}(0)
+2\sum_{l=1}^3\Dtilde_{j-1,k-1,l}(0)+\Dtilde_{j-1,k-1}(0), 
\nl
\Dcomb_{jkl1}(0) &=&
-\sum_{m=1}^3\Dtilde_{j-1,k-1,l-1,m}(0)-\Dtilde_{j-1,k-1,l-1}(0), 
\nl
\Dcomb_{jklm}(0) &=& \Dtilde_{j-1,k-1,l-1,m-1}(0), 
\nnls
\Dcomb^{\fin}_{00001}(0) &=& -\sum_{j=1}^3\Dtilde^{\fin}_{0000j}(0)-D^{\fin}_{0000}(0),\nl
\Dcomb^{\fin}_{0000j}(0) &=& \Dtilde^{\fin}_{0000,j-1}(0), 
\nnls
\Dcomb_{00111}(0) &=& -\sum_{j,k,l=1}^3\Dtilde_{00jkl}(0)
-3\sum_{j,k=1}^3\Dtilde_{00jk}(0)-3\sum_{j=1}^3\Dtilde_{00j}(0)-D_{00}(0),\nl
\Dcomb_{00j11}(0) &=&
\sum_{k,l=1}^3\Dtilde_{00,j-1,kl}(0)+2\sum_{k=1}^3\Dtilde_{00,j-1,k}(0)+\Dtilde_{00,j-1}(0),
\nl
\Dcomb_{00jk1}(0) &=&
-\sum_{l=1}^3\Dtilde_{00,j-1,k-1,l}(0)-\Dtilde_{00,j-1,k-1}(0), 
\nl
\Dcomb_{00jkl}(0) &=& \Dtilde_{00,j-1,k-1,l-1}(0), 
\qquad j,k,l,m=2,3,4.
\eeqar

\section{Consistency checks, applications, and generalizations}
\label{se:checks}

In order to check the explicit results for the 5-point tensor coefficients
we have additionally calculated them by applying the usual Passarino--Veltman
algorithm \cite{Passarino:1978jh}, which is conceptually completely different
from the method of this paper. In the actual comparison we expressed
the redundant terms $E_{00}g^{\mu\nu}$, etc., in the Lorentz decomposition
of $E_{\mu\nu}$, etc., in terms of the coefficients $E_{jk}$, etc., by
exploiting the relation \refeq{gred} for the metric tensor.
The numerical comparison of the coefficients
showed 
agreement between the two methods for non-exceptional
phase-space points.

Moreover, we have investigated the performance of the presented method
in practice. To this end, we have 
implemented the new method into 
the calculation of the one-loop 
QCD corrections to the process $gg/q\bar q\to\Pt\bar\Pt\PH$
\cite{Beenakker:2001rj}
and of the one-loop electroweak corrections to $\Pep\Pem\to\nu_l\bar\nu_l\PH$
\cite{eennH}, which both were originally evaluated using Passarino--Veltman 
reduction. We 
have observed
a drastic improvement in the numerical stability of 
the 5-point function.
While near the phase-space boundary the Passarino--Veltman 
approach could only be rescued by an extrapolation from the inner
phase-space region
(see \citere{Beenakker:2001rj} for details),
this CPU-time-consuming procedure is practically
unnecessary for the approach described above. The results obtained with
the two methods are in mutual agreement within the integration errors,
but owing to the more extensive use of the extrapolation the calculation 
employing the Passarino--Veltman methods takes much 
more
time.

Finally, we consider the possibility of generalizing the method
of this paper to dimensional regularization. The generalization is
trivial in all cases where only abelian soft singularities are involved.
Typical examples are pure QED or electroweak processes with an exactly
massless photon and no 
massless charged particles, or QCD processes that
do not involve external gluons or massless quarks. 
In such cases the soft singularity
arises from diagrams with photon or gluon exchange between two
external (on-shell) massive lines and shows up as a single pole in 
$\epsilon=(4-D)/2$. 
Alternatively, if for $D=4$ an infinitesimal photon or gluon mass
$\lambda$ is chosen as IR regulator, the singularity leads to $\ln\lambda$ 
terms. The correspondence between these regularizations is well known
(see e.g.\ \citere{Catani:2001ef}):
\beq
\ln(\lambda^2) \quad\longleftrightarrow\quad
\frac{(4\pi\mu^2)^{\epsilon}\Gamma(1+\epsilon)}{\epsilon} + \O(\epsilon).
\eeq

If external massless charged particles are involved but no external
massless gluons, \ie if the IR singularities are due to photons or
gluons coupled to massless fermions or sfermions the following
approach can be used.  The whole calculation can be carried out with
mass regulators obeying the hierarchy $\lambda\ll m\ll |Q|$, where
$\lambda$ is again an infinitesimal photon or gluon mass, $m$ is a
small fermion or sfermion mass, and $Q$ denotes a typical mass scale
of the process.  Then these results can be translated into dimensional
regularization as described in \citere{Catani:2001ef} for the complete
QCD and SUSY-QCD amplitudes.

If non-abelian soft singularities or overlapping soft/collinear
singularities are involved the situation is more complicated. 
A convenient possibility to make use of the 4-di\-men\-sio\-nal approach 
of this paper is, for instance, described in \citere{Beenakker:2001rj}.
There, a method is presented for translating $D$-dimensional into 
4-di\-men\-sio\-nal
integrals by constructing regularization-scheme-independent finite
integrals upon subtracting well-defined, simple auxiliary integrals with
the same singularity structure. These auxiliary integrals are
entirely built from 3-point functions. In summary, this means that
$D$-dimensional 5-point integrals are first converted into
4-dimensional 5-point integrals (regularized with masses or off-shell
momenta) 
and 3-point integrals. The 4-dimensional 5-point integrals are then
decomposed into 4-point integrals with the method described 
in this paper.

\section{Conclusion}
\label{se:concl}

A method for reducing one-loop tensor 5-point integrals to related 
standard 4-point integrals is
proposed that entirely avoids inverse
Gram matrices, which are potential sources of numerical instabilities
in practice.  The presented explicit results for tensor coefficients
of 1-, 2-, 3-, 4-, and 5-point functions comprise all expressions
needed to deal with $1\to2$, $2\to2$, and $2\to3$ particle reactions
with up to four external gauge bosons of renormalizable gauge theories
and all expressions, apart from 6-point functions, for $2\to4$
particle reactions with up to 
two
external 
bosons. 
The relevant 5-point functions
are UV finite, but may contain IR (soft or collinear) singularities.
In the explicitly given results a four-dimensional regularization scheme is
assumed, but possible ways of translating them into dimensional regularization
are described.

\appendix
\section*{Appendix}

\section{Calculation of extra terms resulting from UV-divergent
  4-point functions}
\label{app:UVterms}

For the reduction of $E_{\mu\nu\rho}$ and $E_{\mu\nu\rho\si}$ we need
to evaluate the UV-divergent 4-point functions
$D^{(\La)}_{\al\mu\nu\rho}$ and $D^{(\La)}_{\al\mu\nu\rho\si}$ 
introduced in \refse{se:derivation}. Recall that these integrals are
defined in four space--time 
dimensions.
Going to $D$ dimensions and using the large-mass expansion
\cite{Smirnov:1996ng} for the regularization parameter $\La$, we
obtain for these integrals
\beqar
D^{(\La)}_{\al\mu\nu\rho}&=& \left. 
\frac{(2\pi\mu)^{4-D}}{\ri\pi^{2}}\int\! \rd^{D}q\, \frac{q_\al q_{\mu}q_\nu
  q_{\rho}}{N_{0}N_{1}N_{2}N_{3}}\frac{-\La^2}{q^2-\La^2}
  \right|_{D=4}
\nl
&\asymLa& D_{\al\mu\nu\rho} + 
\frac{(2\pi\mu)^{4-D}}{\ri\pi^{2}}\int\! \rd^{D}q\, \frac{q_\al q_{\mu}q_\nu
  q_{\rho}}{(q^2)^4}\frac{-\La^2}{q^2-\La^2}
  + \O(\La^{-2}\ln\La) + \O(D-4) \hspace{2em}
\eeqar
and
\beqar
D^{(\La)}_{\al\mu\nu\rho\si}&=& \left. 
\frac{(2\pi\mu)^{4-D}}{\ri\pi^{2}}\int\! \rd^{D}q\, \frac{q_\al q_{\mu}q_\nu
  q_{\rho}q_\si}{N_{0}N_{1}N_{2} N_{3}}\frac{-\La^2}{q^2-\La^2}
  \right|_{D=4}
\nl
&\asymLa& D_{\al\mu\nu\rho\si} +
\frac{(2\pi\mu)^{4-D}}{\ri\pi^{2}}\int\! \rd^{D}q\, \frac{q_\al q_{\mu}q_\nu
  q_{\rho}q_\si}{(q^2)^4}\frac{-\La^2}{q^2-\La^2}
\left[1-\sum_{l=0}^3\frac{2qp_l}{q^2}\right]
\nl && {}
  + \O(\La^{-2}\ln\La) + \O(D-4),
\eeqar
where $D_{\al\mu\nu\rho}$ and $D_{\al\mu\nu\rho\si}$ are the
usual 4-point functions in dimensional regularization 
as defined in \refeq{CDEint}, 
and we included
a non-vanishing momentum $p_0$ in $N_0$ for later convenience.
The extra vacuum integrals can be evaluated by standard methods;
the results are
\beqar
&&\frac{(2\pi\mu)^{4-D}}{\ri\pi^{2}}\int\! \rd^{D}q\, \frac{q_\al q_{\mu}q_\nu
  q_{\rho}}{(q^2)^4}\frac{-\La^2}{q^2-\La^2}
=-\frac{1}{24}g_{\al[\mu}g_{\nu\rho]}
\left[\Delta+\frac{11}{6}+\ln\frac{\mu^2}{\La^2}\right]+\O(D-4),\nl
&&\frac{(2\pi\mu)^{4-D}}{\ri\pi^{2}}\int\! \rd^{D}q\, \frac{q_\al q_{\mu}q_\nu
  q_{\rho}q_\si}{(q^2)^4}\frac{-\La^2}{q^2-\La^2}
\left[1-\sum_{l=0}^3\frac{2qp_l}{q^2}\right]\nl
&&\qquad=\frac{1}{96}\Bigl(\sum_{l=0}^3 p_l\Bigr){}_{[\al}g_{\mu[\nu}g_{\rho\si]]} 
\left[\Delta+\frac{25}{12}+\ln\frac{\mu^2}{\La^2}\right]+\O(D-4),
\eeqar
where 
\beq
\De = \frac{2}{4-D} -\gamma_{\mathrm{E}} +\ln{4\pi}
\eeq
represents the UV singularities in dimensional regularization,
$\gamma_{\mathrm{E}}$ is Euler's constant, and $\mu$ is the
dimensionful parameter of dimensional regularization.  

Since the UV singularities of $D^{(\La)}_{\mu_1\ldots\nu_P}$ are
regularized by $\La$, the poles in $D-4$ in the vacuum integrals
cancel the poles in $D_{\al\mu\nu\rho}$ and $D_{\al\mu\nu\rho\si}$, and
we can write
\beq
D^{(\La)}_{\mu_1\ldots\nu_P}\asymLa
D^{\fin}_{\mu_1\ldots\nu_P}+\De_{\mu_1\ldots\nu_P}
\eeq
with
\beqar
\De_{\mu\nu\rho\si}&=&
-\frac{1}{24}g_{\mu[\nu}g_{\rho\si]}
\left[\frac{11}{6}+\ln\frac{\mu^2}{\La^2}\right]+\O(D-4),\nl
\De_{\al\mu\nu\rho\si}
&=&\frac{1}{96}\Bigl(\sum_{l=0}^3 p_l\Bigr){}_{[\al}g_{\mu[\nu}g_{\rho\si]]} 
\left[\frac{25}{12}+\ln\frac{\mu^2}{\La^2}\right]+\O(D-4)
\eeqar
and
\beqar
D^{\fin}_{\mu\nu\rho\si}&=&D_{\mu\nu\rho\si}
-\frac{1}{24}g_{\mu[\nu}g_{\rho\si]}\De+\O(D-4),\nl
D^{\fin}_{\al\mu\nu\rho\si}
&=&D_{\al\mu\nu\rho\si}
+\frac{1}{96}\Bigl(\sum_{l=0}^3 p_l\Bigr){}_{[\al}g_{\mu[\nu}g_{\rho\si]]} 
\De+\O(D-4).
\eeqar
The quantities $D^{(\La)}_{\mu_1\ldots\nu_P}$,
$D^{\fin}_{\mu_1\ldots\nu_P}$, and $\De_{\mu_1\ldots\nu_P}$
are finite in $D$ dimensions, so that we can perform
the limit $D\to4$ there.

In the reduction of $E_{\mu\nu\rho}$, UV-divergent contributions
appear only in the difference
$[\De_{\al\mu\nu\rho}(i)-\De_{\al\mu\nu\rho}(0)]$. Since
$\De_{\al\mu\nu\rho}$ is momentum independent, the difference, and thus
$\UVdet_{\mu\nu\rho}$, vanishes.

For $E_{\mu\nu\rho\si}$, we have contributions from
$\De_{\mu\nu\rho\si}(0)$ and
$\De_{\al\mu\nu\rho\si}(i)-\De_{\al\mu\nu\rho\si}(0)$.  
These terms are given by
\beqar\label{Deltares}
\De_{\mu\nu\rho\si}(0)&=&-\frac{1}{24}g_{\mu[\nu}g_{\rho\si]}
 \left[\frac{11}{6}+\ln\frac{\mu^2}{\La^2}\right] + \O(D-4),\nl
\De_{\al\mu\nu\rho\si}(i)-\De_{\al\mu\nu\rho\si}(0) &=&
        \phantom{{}-{}}
  \frac{1}{96}\Bigl(\sum_{l=1\atop l\ne i}^4
  p_l\Bigr){}_{[\al}g_{\mu[\nu}g_{\rho\si]]}
  \left[\frac{25}{12}+\ln\frac{\mu^2}{\La^2}\right]
\nl
&&{}-\frac{1}{96}\Bigl(\sum_{l=1}^4 p_l\Bigr){}_{[\al}g_{\mu[\nu}g_{\rho\si]]}
\left[\frac{25}{12}+\ln\frac{\mu^2}{\La^2}\right] + \O(D-4)\nl
&=& -\frac{1}{96} p_{i,[\al}g_{\mu[\nu}g_{\rho\si]]}
\left[\frac{25}{12}+\ln\frac{\mu^2}{\La^2}\right] + \O(D-4).
\eeqar

Inserting \refeq{Deltares} into \refeq{eq:UVdetdef} and using
\beqar
\sum_{i,j=1}^4 (-1)^{i+j}  \, \det(\Zhat^{(4)}_{ij}) \, 2p_j^\alpha p_{i,\alpha}
= \sum_{i,j=1}^4 \det(Z^{(4)}) \, 
(Z^{(4)})^{-1}_{ij} \, 2p_j^\alpha p_{i,\alpha}
= 4\det(Z^{(4)}),
\eeqar
which employs the relation between $\det(\Zhat^{(4)}_{ij})$ and the
inverse Gram matrix $(Z^{(4)})^{-1}$,
we find
\beqar
\UVdet_{\mu\nu\rho\si}&=&
-\frac{1}{48} g_{[\mu\nu}g_{\rho]\sigma} \det(Z^{(4)}) + \O(D-4).
\eeqar

\section{Reduction of tensor 1-, 2-, 3-, and 4-point functions}
\label{app:1234reduction}

One-loop tensor $N$-point integrals have the general form
\beq
\label{tensorint}
\begin{array}[b]{l}
\hspace{-1em} T^{N}_{\mu_{1}...\mu_{P}}(p_{1},...,p_{N-1},m_{0},...,m_{N-1})=
\displaystyle{\frac{(2\pi\mu)^{4-D}}{i\pi^{2}}\int \rd^{D}q
\frac{q_{\mu_{1}}\cdots q_{\mu_{P}}}
{N_0N_1\ldots N_{N-1}}}
\earr
\eeq
with $N_i$, $i=0,\ldots,N-1$, 
defined in \refeq{D0Di}.
Following the notation of \citere{'tHooft:1978xw}, \ie $T^{1}\to A,$
$T^{2}\to B,$ $T^{3}\to C,$ $T^{4}\to D$, and using the conventions of
\citere{Denner:kt}, we decompose the genuine tensor integrals into 
Lorentz-covariant structures:
\beqar\label{eq:tensdec}
A^\mu &=& \rlap{0,} \hspace{8em}
A^{\mu\nu} = g^{\mu\nu}A_{00}, \nn\\[.2em]
B^\mu &=& \rlap{$ p^\mu_1 B_1, $} \hspace{8em}
B^{\mu\nu} = g^{\mu\nu}B_{00}+p^\mu_1 p^\nu_1 B_{11}, \nn\\[.2em]
B^{\mu\nu\rho} &=& 
g^{[\mu\nu}p_1^{\rho]}B_{001}+p^\mu_1 p^\nu_1 p^\rho_1 B_{111},\nl
C^\mu &=& \rlap{$ \disp\sum_{j=1}^{2} p^\mu_j C_j, $} \hspace{8em}
C^{\mu\nu} = g^{\mu\nu}C_{00}+\sum_{j,k=1}^{2}p^\mu_j p^\nu_k C_{jk}, \nn\\
C^{\mu\nu\rho} &=& \sum_{j=1}^{2} 
g^{[\mu\nu}p_{j}^{\rho]}C_{00j}+
\sum_{j,k,l=1}^2 p_{j}^{\mu}p_{k}^{\nu}p_{l}^{\rho}C_{jkl}, \nn\\
C^{\mu\nu\rho\si} &=& 
g^{[\mu\nu}g^{\rho]\si}C_{0000}
+\sum_{j,k=1}^{2}
(g^{[\mu\nu}p_{j}^{\rho]}p_k^\si+g^{\si[\mu}p_{j}^{\nu}p_k^{\rho]})C_{00jk}
+\sum_{j,k,l,m=1}^2 p_{j}^{\mu}p_{k}^{\nu}p_{l}^{\rho}p_m^\si C_{jklm},\nl
D^\mu &=& \rlap{$ \disp\sum_{j=1}^{3} p^\mu_j D_j, $} \hspace{8em}
D^{\mu\nu} = g^{\mu\nu}D_{00}+\sum_{j,k=1}^{3}p^\mu_j p^\nu_k D_{jk}, \nn\\
D^{\mu\nu\rho} &=& \sum_{j=1}^{3}g^{[\mu\nu}p_{j}^{\rho]}D_{00j}
+\sum_{j,k,l=1}^3 p_{j}^{\mu}p_{k}^{\nu}p_{l}^{\rho}D_{jkl},\nl
D^{\mu\nu\rho\si} &=& 
g^{[\mu\nu}g^{\rho]\si}D_{0000}
+\sum_{j,k=1}^{3}
(g^{[\mu\nu}p_{j}^{\rho]}p_k^\si+g^{\si[\mu}p_{j}^{\nu}p_k^{\rho]}
)D_{00jk}
+\sum_{j,k,l,m=1}^3 p_{j}^{\mu}p_{k}^{\nu}p_{l}^{\rho}p_m^\si D_{jklm},\nl
D^{\mu\nu\rho\si\tau} &=& \sum_{j=1}^{3} 
g^{[[\mu\nu}g^{\rho]\si}p_j^{\tau]}
D_{0000j}
+\sum_{j,k,l,=1}^{3}
(g^{[\mu\nu}p_{j}^{\rho}p_k^{\si} p_l^{\tau]}+g^{[\mu\rho}p_{j}^{\tau}p_k^{\nu} p_l^{\si]})D_{00jkl}
\nl&&{}
+\sum_{j,k,l,m,n=1}^3 p_{j}^{\mu}p_{k}^{\nu}p_{l}^{\rho}p_m^\si p_n^\tau D_{jklmn}.
\label{decomptensor}
\eeqar
Because of the symmetry of the tensor $T^{N}_{\mu_{1}...\mu_{P}}$ all
coefficients $B_1$,...,$D_{jklmn}$ are symmetric under permutation of 
all indices. They are reduced to scalar integrals
recursively by inversion of systems of linear equations. The inhomogeneity 
of 
these 
equations
consists of coefficients of lower rank
\cite{Passarino:1978jh}. The equations of this system are obtained by
contracting one integration momentum $q_{\mu_1}$ with the $(N-1)$ external
momenta $p_i^{\mu_1}$ and for $N\ge 2$ also by contraction with the metric
$g^{\mu_1\mu_2}$. Using 
\beq
2p_i q = N_i-N_0-f_i \quad \mbox{with}
\quad f_i=p_i^2-m_i^2+m_0^2,
\label{contrtensor}
\eeq
in (\ref{tensorint}), the first two terms on the right-hand side of 
(\ref{contrtensor}) each cancel exactly one propagator denominator of
$q^{\mu_{1}}T^{N}_{\mu_{1}...\mu_{P}}$, the third term is proportional to
$T^{N}_{\mu_{2}...\mu_{P}}$. 
Likewise the contraction with $g^{\mu_1\mu_2}$ yields a factor $q^2$
in the numerator of $g^{\mu_1\mu_2}T^{N}_{\mu_{1}...\mu_{P}}$, which
can be written as $q^2 = N_0 + m_0^2$.
The $N_0$ term cancels the first propagator, the second term leads to the 
tensor $T^{N}_{\mu_{3}...\mu_{P}}$. With the abbreviation 
$T^{N-1}_{\mu_{1}...\mu_{P}}(i)$,
denoting 
$T^{N}_{\mu_{1}...\mu_{P}}$ 
with the $i$th denominator omitted, this yields
\beqar
2p_i^{\mu_{1}}T^{N}_{\mu_{1}...\mu_{P}} &=&
T^{N-1}_{\mu_{2}...\mu_{P}}(i)-T^{N-1}_{\mu_{2}...\mu_{P}}(0)-
f_i T^{N}_{\mu_{2}...\mu_{P}}, \nn\\[.2em]
g^{\mu_1\mu_2}T^{N}_{\mu_{1}\mu_{2}...\mu_{P}} &=&
T^{N-1}_{\mu_{3}...\mu_{P}}(0)+m_0^2T^{N}_{\mu_{3}...\mu_{P}}.
\label{recurrtensor}\eeqar
Note that for $T^{N-1}_{\mu_{1}...\mu_{P}}(0)$ a shift of the integration
momentum $q^\mu\to q^\mu-p_1^\mu$ has to be done in order to achieve the 
standard form (\ref{tensorint}). The tensor integrals with shifted
momenta $\tilde T^{N-1}_{\mu_{1}...\mu_{P}}(0)$ are defined as in
\refeq{Dshifted}. Expressing the tensors in (\ref{recurrtensor}) by
(\ref{decomptensor}), the desired recurrence relations can be read off
by comparing coefficients.  

In the following we summarize the results for the reduction of all
tensor coefficients defined in \refeq{eq:tensdec}. These results can
also be read off from the generic results in Section 4.2 of
\citere{Denner:kt}. Note that several tensor coefficients can be obtained from
different reduction formulas. This allows to check the correctness and
the numerical stability of the results.

The tensor coefficient of the relevant 1-point function reads
\beq
A_{00} = \frac{1}{4}m_0^2 A_0 + \frac{1}{8}m_0^4.
\eeq
For the tensor coefficients of 2-point functions we find
\beqar
B_1  &=& \frac{1}{2p_1^2}\left[A_0(1)-A_0(0)-f_1 B_0\right], \\[.3em]
B_{00} &=& \frac{1}{6}\left[A_0(0)+f_1 B_1+2m_0^2 B_0+m_0^2+m_1^2
-\textfrac{1}{3}p_1^2\right], \nn\\[.1em]
B_{11} &=& \frac{1}{3p_1^2}\left[A_0(0)-2f_1 B_1-m_0^2 B_0
-\textfrac{1}{2}\left(m_0^2+m_1^2-\textfrac{1}{3}p_1^2\right)\right], \\
B_{001} &=& \frac{1}{2p_1^2}\left[ A_{00}(1)-A_{00}(0)-f_1 B_{00}
\right]\nl
        &=& \frac{1}{8}\left[2m_0^2B_1-A_0(0)+f_1 B_{11}
-\textfrac{1}{6}(2m_0^2+4m_1^2-p_1^2)\right],\nl
B_{111}
        &=& -\frac{1}{4p_1^2}\left[A_0(0)+3f_1 B_{11}+2m_0^2 B_1
-\textfrac{1}{6}\left(2m_0^2+4m_1^2-p_1^2\right)\right].
\eeqar
For the 3-point functions we obtain for the vector case
\beqar
C_{i} &=& \sum_{n=1}^2(Z^{(2)})^{-1}_{in}R^1_{n}, \qquad i=1,2,
\qquad
Z^{(2)}=\left(
\barr{c@{\:\:\:}c}
 2p_{1}p_1 & 2p_{1}p_{2} \\
 2p_{2}p_1 & 2p_{2}p_{2} 
\earr
\right),
\eeqar
with
\beqar
R^1_{n}&=& B_{0}(n)-B_{0}(0) - f_n C_{0}, \qquad n=1,2,
\eeqar
for the second-rank tensor case
\beqar
C_{00} &=& \textfrac{1}{2}m_0^2C_{0}+\textfrac{1}{4}B_{0}(0)
+\textfrac{1}{4}\sum_{m=1}^2f_m C_{m}+\textfrac{1}{4},
\nnls
C_{ij} &=& \sum_{n=1}^2(Z^{(2)})^{-1}_{in}[R^2_{nj}-2C_{00}\de_{nj}]
, \qquad i,j=1,2,
\eeqar
with [$i_n$ is defined in \refeq{eq:defji}]
\beqar
R^2_{ni}&=& B_{1}(n)\debar_{ni}-\Bcomb_{i}(0) - f_n C_{i}, \qquad n,i=1,2,
\eeqar
and
\beqar
\Bcomb_{1}(0) &=& -\Btilde_{1}(0)-B_{0}(0), \qquad
\Bcomb_{2}(0) = \Btilde_{1}(0),
\eeqar
for the third-rank tensor case
\beqar
C_{00i} &=& \textfrac{1}{3}m_0^2C_{i}+\textfrac{1}{6}\Bcomb_{i}(0)
+\textfrac{1}{6}\sum_{j=1}^2 f_j C_{ij}-\textfrac{1}{18},
\nnls
C_{00i} &=& \sum_{n=1}^2(Z^{(2)})^{-1}_{in}R^3_{n00}, 
\nnls
C_{ijk} &=&
\sum_{n=1}^2(Z^{(2)})^{-1}_{in}[R^3_{njk}-2C_{00j}\de_{nk}-2C_{00k}\de_{nj}]
, \qquad i,j,k=1,2,
\eeqar
with
\beqar
R^3_{n00}&=& B_{00}(n)-B_{00}(0) - f_n C_{00}, 
\nnls
R^3_{nij}&=& B_{11}(n)\debar_{ni}\debar_{nj}
-\Bcomb_{ij}(0) - f_n C_{ij}, \qquad n,i,j=1,2,
\eeqar
and
\beqar
\Bcomb_{11}(0) &=& \Btilde_{11}(0)+2\Btilde_{1}(0)+B_{0}(0),\nl 
\Bcomb_{12}(0) &=& -\Btilde_{11}(0)-\Btilde_{1}(0), \nl
\Bcomb_{22}(0) &=& \Btilde_{11}(0),
\eeqar
and for the fourth-rank tensor case
\beqar
C_{0000} &=& \textfrac{1}{4}m_0^2C_{00}+\textfrac{1}{8}B_{00}(0)
+\textfrac{1}{8}\sum_{i=1}^2f_i C_{00i} +\textfrac{1}{48}(m_0^2+m_1^2+m_2^2)
-\textfrac{1}{192}[p_1^2+p_2^2+(p_1-p_2)^2],
\nnls
C_{00ij} &=& \textfrac{1}{4}m_0^2C_{ij}+\textfrac{1}{8}\Bcomb_{ij}(0)
+\textfrac{1}{8}\sum_{k=1}^2 f_k C_{ijk}+\textfrac{1}{96}(1+\de_{ij}),
\nnls
C_{00ij} &=&
\sum_{n=1}^2(Z^{(2)})^{-1}_{in}[R^4_{n00j}-2C_{0000}\de_{nj}],
\nnls
C_{ijkl} &=& \sum_{n=1}^2(Z^{(2)})^{-1}_{in}[R^4_{njkl}-2C_{00jk}\de_{nl}
-2C_{00kl}\de_{nj}-2C_{00lj}\de_{nk}],\qquad i,j,k,l=1,2,
\eeqar
with
\beqar
R^4_{n00i}&=& B_{001}(n)\debar_{ni}-\Bcomb_{00i}(0) - f_n C_{00i},
\nnls
R^4_{nijk}&=&  B_{111}(n)\debar_{ni}\debar_{nj}\debar_{nk}
-\Bcomb_{ijk}(0) - f_n C_{ijk},\qquad n,i,j,k=1,2,
\eeqar
and
\beqar
\Bcomb_{001}(0) &=& -\Btilde_{001}(0)-B_{00}(0),  \qquad
\Bcomb_{002}(0) = \Btilde_{001}(0), 
\nnls
\Bcomb_{111}(0) &=& -\Btilde_{111}(0)-3\Btilde_{11}(0)
-3\Btilde_{1}(0)-B_{0}(0),\nl
\Bcomb_{112}(0) &=& \Btilde_{111}(0)+2\Btilde_{11}(0)+\Btilde_{1}(0),\nl
\Bcomb_{122}(0) &=& -\Btilde_{111}(0)-\Btilde_{11}(0), \nl
\Bcomb_{222}(0) &=& \Btilde_{111}(0).
\eeqar

The tensor coefficients of 4-point functions are given by
\beqar
D_{i} &=& \sum_{n=1}^3(Z^{(3)})^{-1}_{in}S^1_{n}, \qquad i=1,2,3,
\qquad
Z^{(3)}=\left(
\barr{c@{\:\:\:}c@{\:\:\:}c}
 2p_{1}p_1 & 2p_{1}p_{2} & 2p_{1}p_{3} \\
 2p_{2}p_1 & 2p_{2}p_{2} & 2p_{2}p_{3} \\
 2p_{3}p_1 & 2p_{3}p_{2} & 2p_{3}p_{3} 
\earr
\right),
\eeqar
with
\beqar
S^1_{n}&=& C_{0}(n)-C_{0}(0) - f_n D_{0}, \qquad n=1,2,3,
\eeqar
for the vector case, by
\beqar
D_{00} &=& m_0^2D_{0}+\textfrac{1}{2}C_{0}(0)
+\textfrac{1}{2}\sum_{m=1}^3f_m D_{m},
\nnls
D_{ij} &=& \sum_{n=1}^3(Z^{(3)})^{-1}_{in}[S^2_{nj}-2D_{00}\de_{nj}]
, \qquad i,j=1,2,3,
\eeqar
with
\beqar
S^2_{ni}&=&  C_{i_n}(n)\debar_{ni}-\Ccomb_{i}(0) - f_n D_{i}, \qquad n,i=1,2,3,
\eeqar
and
\beqar
\Ccomb_{1}(0) &=& -\sum_{i=1}^2 \Ctilde_{i}(0)-C_{0}(0),\qquad
\Ccomb_{i}(0) = \Ctilde_{i-1}(0), \qquad i=2,3,
\eeqar
for the second-rank tensor case, by
\beqar
D_{00i} &=& \textfrac{1}{2}m_0^2D_{i}+\textfrac{1}{4}\Ccomb_{i}(0)
+\textfrac{1}{4}\sum_{j=1}^3 f_j D_{ij},
\nnls
D_{00i} &=& \sum_{n=1}^3(Z^{(3)})^{-1}_{in}S^3_{n00}, 
\nnls
D_{ijk} &=&
\sum_{n=1}^3(Z^{(3)})^{-1}_{in}[S^3_{njk}-2D_{00j}\de_{nk}-2D_{00k}\de_{nj}]
, \qquad i,j,k=1,2,3,
\eeqar
with
\beqar
S^3_{n00}&=& C_{00}(n)-C_{00}(0) - f_n D_{00}, 
\nnls
S^3_{nij}&=&  C_{i_nj_n}(n)\debar_{ni}\debar_{nj}-\Ccomb_{ij}(0) - f_n D_{ij}, \qquad n,i,j=1,2,3,
\eeqar
and
\beqar
\Ccomb_{11}(0) &=& \sum_{i,j=1}^2\Ctilde_{ij}(0)
+2\sum_{i=1}^2\Ctilde_{i}(0)+C_{0}(0), \nl
\Ccomb_{1i}(0) &=& -\sum_{j=1}^2\Ctilde_{i-1,j}(0)-
\Ctilde_{i-1}(0), 
\nl
\Ccomb_{ij}(0) &=& \Ctilde_{i-1,j-1}(0), \qquad i,j=2,3,
\eeqar
for the third-rank tensor case, by
\beqar
D_{0000} &=& \textfrac{1}{3}m_0^2D_{00}+\textfrac{1}{6}C_{00}(0)
+\textfrac{1}{6}\sum_{i=1}^3f_i D_{00i} +\textfrac{1}{36},
\nnls
D_{00ij} &=& \textfrac{1}{3}m_0^2D_{ij}+\textfrac{1}{6}\Ccomb_{ij}(0)
+\textfrac{1}{6}\sum_{k=1}^3 f_k D_{ijk},
\nnls
D_{00ij} &=& \sum_{n=1}^3(Z^{(3)})^{-1}_{in}[S^4_{n00j}-2D_{0000}\de_{nj}],
\nnls
D_{ijkl} &=& \sum_{n=1}^3(Z^{(3)})^{-1}_{in}[S^4_{njkl}-2D_{00jk}\de_{nl}
-2D_{00kl}\de_{nj}-2D_{00lj}\de_{nk}],\qquad i,j,k,l=1,2,3,\nln
\eeqar
with
\beqar
S^4_{n00i}&=&  C_{00i_n}(n)\debar_{ni}-\Ccomb_{00i}(0) - f_n D_{00i},
\nnls
S^4_{nijk}&=&  C_{i_n j_n k_n}(n)\debar_{ni}\debar_{nj}\debar_{nk}
-\Ccomb_{ijk}(0) - f_n D_{ijk},\qquad n,i,j,k=1,2,3,
\eeqar
and
\beqar
\Ccomb_{001}(0) &=& -\sum_{i=1}^2 \Ctilde_{00i}(0)-C_{00}(0),
\qquad
\Ccomb_{00i}(0) = \Ctilde_{00,i-1}(0), 
\nnls
\Ccomb_{111}(0) &=&
-\sum_{i,j,k=1}^2\Ctilde_{ijk}(0)-3\sum_{i,j=1}^2\Ctilde_{ij}(0)
-3\sum_{i=1}^2\Ctilde_{i}(0)-C_{0}(0),\nl
\Ccomb_{11i}(0) &=& \sum_{j,k=1}^2\Ctilde_{i-1,jk}(0)
+2\sum_{j=1}^2\Ctilde_{i-1,j}(0)+\Ctilde_{i-1}(0), 
\nl
\Ccomb_{1ij}(0) &=& -\sum_{k=1}^2 \Ctilde_{i-1,j-1,k}(0)
-\Ctilde_{i-1,j-1}(0), 
\nl
\Ccomb_{ijk}(0) &=& \Ctilde_{i-1,j-1,k-1}(0), \qquad i,j,k=2,3,
\eeqar
for the fourth-rank tensor case, and by
\beqar
D_{0000i} &=& \textfrac{1}{4}m_0^2D_{00i}+\textfrac{1}{8}\Ccomb_{00i}(0)
+\textfrac{1}{8}\sum_{j=1}^3 f_j D_{00ij} -\textfrac{1}{192},
\nnls
D_{00ijk} &=& \textfrac{1}{4}m_0^2D_{ijk}+\textfrac{1}{8}\Ccomb_{ijk}(0)
+\textfrac{1}{8}\sum_{m=1}^3 f_m D_{ijkm},
\nnls
D_{0000i} &=& \sum_{n=1}^3(Z^{(3)})^{-1}_{in}S^5_{n0000},
\nnls
D_{00ijk} &=&
\sum_{n=1}^3(Z^{(3)})^{-1}_{in}[S^5_{n00jk}-2D_{0000j}\de_{nk}
-2D_{0000k}\de_{nj}],
\nnls
D_{ijklm} &=&
\sum_{n=1}^3(Z^{(3)})^{-1}_{in}[S^5_{njklm}-2D_{00jkl}\de_{nm}
-2D_{00klm}\de_{nj}-2D_{00lmj}\de_{nk}-2D_{00mjk}\de_{nl}],\nl
\qquad &&i,j,k,l,m=1,2,3,
\eeqar
with
\beqar
S^5_{n0000}&=& C_{0000}(n)-C_{0000}(0) - f_n D_{0000},
\nnls
S^5_{n00ij}&=&  C_{00i_nj_n}(n)\debar_{ni}\debar_{nj}
-\Ccomb_{00ij}(0) - f_n D_{00ij},
\nnls
S^5_{nijkl}&=&
C_{i_nj_nk_nl_n}(n)\debar_{ni}\debar_{nj}\debar_{nk}\debar_{nl}
-\Ccomb_{ijkl}(0) - f_n D_{ijkl},\qquad n,i,j,k,l=1,2,3,
\hspace{2em}
\eeqar
and
\beqar
\Ccomb_{0011}(0) &=& \sum_{i,j=1}^2 \Ctilde_{00ij}(0)
+2\sum_{i=1}^2\Ctilde_{00i}(0)+C_{00}(0),\nl
\Ccomb_{00i1}(0) &=& -\sum_{j=1}^2 \Ctilde_{00,i-1,j}(0)
-\Ctilde_{00,i-1}(0),
\nl
\Ccomb_{00ij}(0) &=& \Ctilde_{00,i-1,j-1}(0), 
\nnls
\Ccomb_{1111}(0) &=&
\sum_{i,j,k,l=1}^2\Ctilde_{ijkl}(0)+4\sum_{i,j,k=1}^2\Ctilde_{ijk}(0)
+6\sum_{i,j=1}^2\Ctilde_{ij}(0)+4\sum_{i=1}^2\Ctilde_{i}(0)
+C_{0}(0),
\nl
\Ccomb_{111i}(0) &=&
-\sum_{j,k,l=1}^2\Ctilde_{i-1,jkl}(0)-3\sum_{j,k=1}^2\Ctilde_{i-1,jk}(0)
-3\sum_{j=1}^2\Ctilde_{i-1,j}(0)-\Ctilde_{i-1}(0), \nl
\Ccomb_{11ij}(0) &=& \sum_{k,l=1}^2\Ctilde_{i-1,j-1,kl}(0)
+2\sum_{k=1}^2\Ctilde_{i-1,j-1,k}(0)+\Ctilde_{i-1,j-1}(0), 
\nl
\Ccomb_{1ijk}(0) &=& -\sum_{l=1}^2 \Ctilde_{i-1,j-1,k-1,l}(0)
-\Ctilde_{i-1,j-1,k-1}(0), 
\nl
\Ccomb_{ijkl}(0) &=& \Ctilde_{i-1,j-1,k-1,l-1}(0), \qquad i,j,k,l=2,3,
\eeqar
for the fifth-rank tensor case.

\section{UV-divergent parts of tensor integrals}
\label{sectendiv}

For practical calculations it is useful to know the UV-divergent parts
of the tensor integrals explicitly. We directly 
give the products of
$D-4$ with all divergent one-loop tensor coefficient integrals appearing in
renormalizable theories up to terms of the order $\O(D-4)$
\beqar\label{TNdiv}
(D-4)\,A_{0}(m_0) &=& -2m_0^{2} ,\nl[1ex]
(D-4)\,A_{00}(m_0) &=& -\textfrac{1}{2}m_0^{4} ,\nl[1ex]
(D-4)\,B_{0}(p_{1},m_{0},m_{1}) &=& -2 ,\nl[1ex]
(D-4)\,B_{1}(p_{1},m_{0},m_{1}) &=& \phantom{-} 1 ,\nl[1ex]
(D-4)\,B_{00}(p_{1},m_{0},m_{1}) &=&
\phantom{-}\textfrac{1}{6}(p_{1}^{2}-3m_{0}^{2}-3m_{1}^{2}) ,\nl[1ex]
(D-4)\,B_{11}(p_{1},m_{0},m_{1}) &=&  -\textfrac{2}{3} ,\nl[1ex]
(D-4)\,B_{001}(p_{1},m_{0},m_{1}) &=&  -\textfrac{1}{12}(p_{1}^{2}-2m_{0}^{2}-4m_{1}^{2}) ,\nl[1ex]
(D-4)\,B_{111}(p_{1},m_{0},m_{1}) &=&  \phantom{-}\textfrac{1}{2} ,\nl[1ex]
(D-4)\,C_{00}(p_{1},p_{2},m_{0},m_{1},m_{2}) &=&  -\textfrac{1}{2} ,\nl[1ex]
(D-4)\,C_{00i}(p_{1},p_{2},m_{0},m_{1},m_{2}) &=&  \phantom{-}\textfrac{1}{6} ,
\nl[1ex]
(D-4)\,C_{0000}(p_{1},p_{2},m_{0},m_{1},m_{2}) &=&
\phantom{-}\textfrac{1}{48}[(p_1-p_2)^2+p_1^2+p_2^2]\nl
&&{}-\textfrac{1}{12}(m_0^2+m_1^2+m_2^2) ,
\nl[1ex]
(D-4)\,C_{00ij}(p_{1},p_{2},m_{0},m_{1},m_{2}) &=&  -\textfrac{1}{24}(1+\de_{ij}) ,
\nl[1ex]
(D-4)\,D_{0000}(p_{1},p_{2},p_{3},m_{0},m_{1},m_{2},m_{3}) &=&
-\textfrac{1}{12} ,
\nl[1ex]
(D-4)\,D_{0000i}(p_{1},p_{2},p_{3},m_{0},m_{1},m_{2},m_{3}) &=&
\phantom{-}\textfrac{1}{48} .
\eeqar
All other scalar coefficients defined in (\ref{eq:tensdec}) are UV finite.

\end{document}